\documentclass[12pt]{iopart}
\usepackage{iopams}  
\expandafter\let\csname equation*\endcsname=\relax 
\expandafter\let\csname endequation*\endcsname=\relax 
\usepackage{amsmath,amssymb,amsthm} 
\usepackage{physics}
\usepackage{bm}
\usepackage{cite}
\usepackage[dvipdfmx]{graphicx}
\usepackage{xparse}
\usepackage{xcolor}
\usepackage{comment}
\usepackage[normalem]{ulem}
\usepackage{cancel}
\usepackage{upgreek}
\usepackage{hyperref}

\def\HFI{H_{\mathrm{FI}}}
\def\Hsys{H_{\mathrm{sys}}}
\def\Hex{H_{\mathrm{ex}}}
\def\la{\langle}
\def\ra{\rangle}
\def\vk{\vb*{k}}
\def\vq{\vb*{q}}
\def\vr{\vb*{r}}
\def\vS{\vb*{S}}

\def\HZ{H_{\mathrm{Z}}}
\def\HT{H_{\mathrm{T}}}
\def\hdc{h_{\mathrm{dc}}}
\def\hac{h_{\mathrm{ac}}}
\def\IS{I_{\rm s}}
\def\kbt{k_{\mathrm{B}}T}

\begin{document}

\title[Spin pumping into two-dimensional systems]{Spin pumping into two-dimensional systems}

\author{Yuya Ominato$^1$, Masaki Yama$^2$, Ai Yamakage$^3$, Mamoru Matsuo$^{4,5,6,7}$
Takeo Kato$^2$}

\address{$^1$ Waseda Institute for Advanced Study, Waseda University, Shinjuku-ku, Tokyo 169-0051, Japan}
\address{$^2$ Institute for Solid State Physics, The University of Tokyo, Kashiwa 277-8581, Japan}
\address{$^3$ Department of Physics, Nagoya University, Nagoya 464-8602, Japan}
\address{$^4$ Kavli Institute for Theoretical Sciences, University of Chinese Academy of Sciences, Beijing, 100190, China}
\address{$^5$ CAS Center for Excellence in Topological Quantum Computation, University of Chinese Academy of Sciences, Beijing 100190, China}
\address{$^6$ Advanced Science Research Center, Japan Atomic Energy Agency, Tokai, 319-1195, Japan}
\address{$^7$ RIKEN Center for Emergent Matter Science (CEMS), Wako, Saitama 351-0198, Japan}

\ead{kato@issp.u-tokyo.ac.jp}
\vspace{10pt}

\begin{indented}
\item[]October 2025
\end{indented}

\begin{abstract}
In this review, we present recent theoretical developments on spin transport phenomena probed by ferromagnetic resonance (FMR) modulation in two-dimensional systems coupled to magnetic materials. We first address FMR linewidth enhancements induced by spin pumping at interfaces, emphasizing their potential as sensitive probes of superconducting pairing symmetries in two-dimensional superconductors. We then examine FMR modulation due to spin pumping into two-dimensional electron gases formed in semiconductor heterostructures, where the interplay of Rashba and Dresselhaus spin–orbit interactions enables gate-controlled spin transport and persistent spin textures. Finally, we investigate spin pumping in monolayer transition-metal dichalcogenides, where spin-valley coupling and Berry curvature effects lead to valley-selective spin excitations. These developments demonstrate that the spin pumping technique provides a versatile tool for probing spin transport and spin-dependent phenomena in low-dimensional systems, offering a basis for future spintronics applications.
\end{abstract}

\section{Introduction}
\label{Sec_intro}

The precise control and detection of spin in solids has been a central topic in the pursuit of novel electronic functionalities. The discovery of anisotropic magnetoresistance \cite{Baibich1988,Binasch1989,Bhatti2017} and the advent of GMR-based magnetic read heads \cite{Dieny1991} revolutionized information storage technology and laid the foundation for the field of spintronics, in which electronic devices exploit not only the charge but also the spin of electrons. In parallel, extensive efforts have been devoted to developing methods for generating, transporting, and detecting spin currents \cite{Johnson1985,Maekawa2023}. More recently, spin transport in nonmagnetic systems \cite{Zutic2004} and spin currents driven by spin–orbit interactions \cite{Hirsch1999,Kato2004,Wunderlich2005} have emerged as major research directions. 

Low-dimensional systems have become important platforms for exploring spin transport phenomena. 
Two-dimensional superconductors exhibit striking properties such as high transition temperatures and large upper critical fields \cite{Saito2017}. 
The spin transport driven by their distinctive pair potential is therefore of interest.
Two-dimensional electron gases (2DEGs) in semiconductor heterostructures, such as GaAs/AlGaAs interfaces \cite{Dingle1978,Ando1982}, provide a controllable environment for studying spin–orbit interactions \cite{Dresselhaus1955,Bychkov1984,Nitta1997,Winkler2003} and their impact on spin transport. Atomically thin materials, such as graphene \cite{Novoselov2004,Novoselov2005} and transition-metal dichalcogenides (TMDCs) \cite{Mak2010}, provide versatile platforms, with TMDCs exhibiting strong spin–orbit coupling and broken inversion symmetry \cite{Xiao2012}. These systems enable access to spin-related phenomena beyond those available in bulk materials, and their compatibility with device integration highlights their potential for scalable spintronics technologies. In this review article, we focus on spin transport in 2DEGs and atomically thin materials, emphasizing the roles of anisotropic superconductivity~\cite{Sigrist1991}, spin–orbit interactions~\cite{Bercioux2015,Manchon2015}, spin–valley coupling \cite{Xiao2012}, and Berry curvature effects \cite{Xiao2010}.

\begin{figure}
\begin{center}
\includegraphics[width=1\hsize]{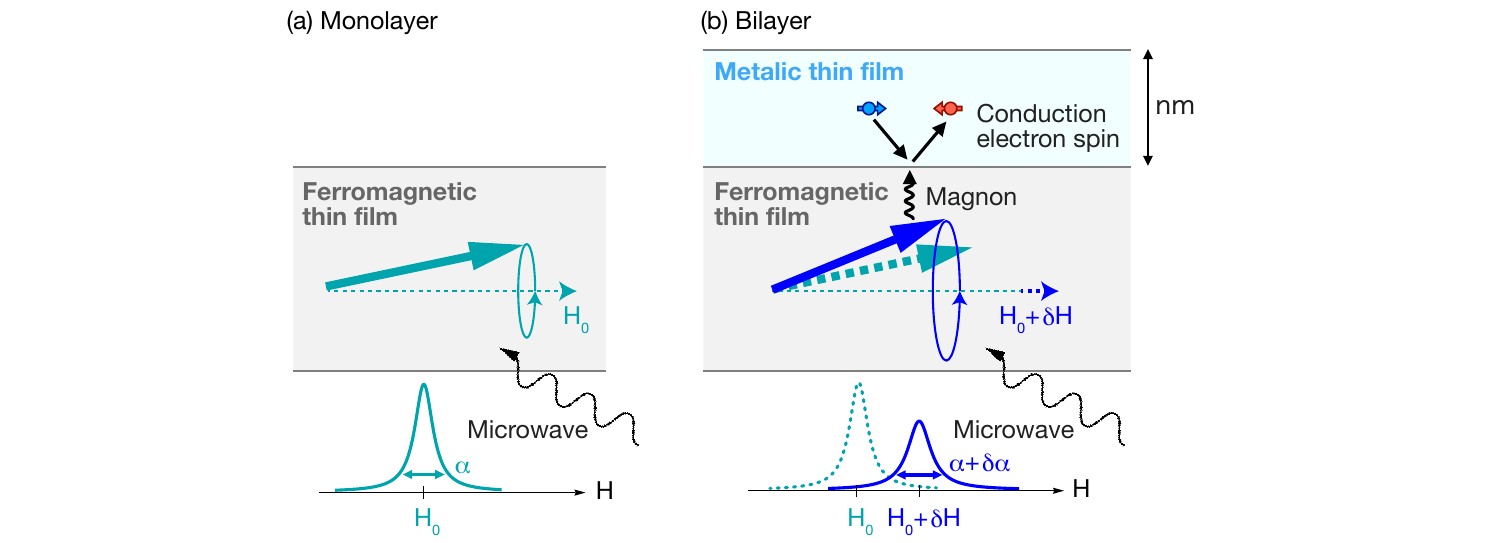}
\end{center}
\caption{A schematic illustration of spin pumping in a paramagnetic metal/ferromagnet bilayer system. Ferromagnetic resonance is induced by microwave irradiation, leading to the injection of a spin current into the metal via interfacial spin exchange interactions. As a back-action of spin pumping, the spin dynamics of the ferromagnet is modulated. This modulation encodes information about the spin excitation of conduction electrons, allowing the determination of their dynamic spin susceptibility in the adjacent metal. Simultaneously, a magnon carrying angular momentum $-\hbar$ is excited in the magnetic system, flipping the conduction electron spin from $+\hbar/2$ to $-\hbar/2$ via interfacial interaction. This process serves as a universal mechanism for the generation of spin currents.}
\label{fig_FMR_signal}
\end{figure}

The exploration of spin transport in low-dimensional systems necessitates sensitive techniques for detecting spin dynamics. One such method is spin pumping, which emerged as a versatile technique for generating spin currents in the history of spintronics \cite{Saitoh2006}.
In this method, a ferromagnet is placed in contact with a nonmagnetic material, and ferromagnetic resonance (FMR) generates a pure spin current in the nonmagnet via interfacial spin exchange interactions \cite{Tserkovnyak2002,Tserkovnyak2005}. Because of its broad applicability to various nonmagnetic materials, spin pumping has become a widely utilized technique \cite{Ando2011}. Simultaneously, spin injection leads to a back-action on the FMR signal, with the resulting FMR modulation reflecting spin excitations in the adjacent nonmagnet \cite{Han2020}. Thus, FMR serves as a spin probe, as shown in Fig.~\ref{fig_FMR_signal}. 
In contrast to conventional spin probes such as nuclear magnetic resonance \cite{Leggett1975} and neutron scattering \cite{Shull1963,Shull1963b,Shull1966}, which often suffer from insufficient sensitivity to atomic layer materials and interface spin dynamics, FMR offers a complementary approach for detecting spin excitations in low-dimensional systems.

In this review, we briefly summarize recent theoretical studies on spin transport phenomena in two-dimensional electron systems, emphasizing the effectiveness of spin pumping as a tool for spin generation and detection of spin excitations.
We review theoretical methods for describing spin pumping in Sec.~\ref{sec_recent_experiment}.
For illustrative examples, we introduce recent theories on spin pumping into two-dimensional superconductors (Sec.~\ref{sec:superconductors}), 2DEGs (Sec.~\ref{sec_SP_2DEG}), and TMDCs (Sec.~\ref{sec_SP_TMDC}).
Finally, we present other applications and summarize this review in Sec.~\ref{sec:perspectives} and Sec.~\ref{sec_summary}, respectively.

\section{Spin Pumping}
\label{sec_recent_experiment}

\subsection{Ferromagnetic Resonance (FMR)}

In a ferromagnet under an external magnetic field, the magnetization within the ferromagnet undergoes the Larmor precession around the effective magnetic field $\bm{H}_{\rm eff}$, which is usually the sum of an external magnetic field and an effective field induced from magnetic anisotropy.
When microwaves are applied to this ferromagnet, the microwave absorption intensity increases significantly if its frequency matches the Larmor precession frequency. This phenomenon is known as ferromagnetic resonance (FMR)~\cite{Vonsovskii1966}. 

Let us consider the case that an oscillating magnetic field ${\bm H}_{\rm ac}(t) =(h_{\rm ac} \cos\omega t,-h_{\rm ac}\sin\omega t,0)$ is applied to a ferromagnet under a static magnetic field ${\bm H}_{\rm eff} = (0,0,-h_{\rm dc})$.
The dynamics of the total spin $\bm{S}_{\rm tot}$ is described by the Landau-Lifshitz-Gilbert (LLG) equation~\cite{Gilbert2004}.
For the spin direction ${\bm m} = {\bm S}_{\rm tot}/S_{\rm tot}$, the LLG equation is written as
\begin{align}
    \frac{d\bm{m}}{dt}=
    \gamma_{\rm g}\bm{m}\times (\bm{H}_{\rm eff}+{\bm H}_{\rm ac}(t))
    -
    \alpha_{\rm G}\bm{m}\times\frac{d\bm{m}}{dt}.\label{eq:LLG}
\end{align}
Here, $\gamma_{\rm g}$ ($<0$) is a gyromagnetic ratio and $\alpha_{\rm G}$ is the Gilbert damping constant.
For a small amplitude of precessional motion around the $z$-axis ($m_x,m_y \ll m_z$), the spin direction oscillates around a stable solution as ${\bm m}(t) = {\bm m}_0 + \delta {\bm m}(t)$.
The time-dependent modulation is described by the LLG equation as
\begin{align}
(1+i\alpha_{\rm G})\frac{dm_+}{dt}=-i\omega_{\bm 0} m_+ + i\gamma_{\rm g}h_{\rm ac} e^{-i\omega t}.\label{eq:Mhh}
\end{align}
where $m_+=\delta m_x+i\delta m_y$, and $\omega_{\bm 0} = |\gamma_{\rm g}| h_{\rm dc}$ is the resonance frequency.
The linear response coefficient $\chi_{\rm FI} =S_{\rm tot} m_+/(-h_{\rm ac} e^{-i\omega t})$ is calculated as
\begin{align}
\chi_{\rm FI} &= \frac{|\gamma_{\rm g}|S_{\rm tot}}{\omega-\omega_{\bm 0}+i\alpha_{\rm G}\omega}.
\label{chiLLG}
\end{align}
The microwave absorption is proportional to the imaginary part of the above response function, leading to the Lorentzian form of the resonance peak (see Fig.~\ref{fig_FMR_signal}(a)).

\subsection{Spin mixing conductance}
\label{sec:SPSMC}

Next, let us consider a junction composed of a ferromagnet and a normal metal (NM) (see Fig.~\ref{fig_FMR_signal}(b)).
In addition to the Gilbert damping, the interfacial exchange coupling produces additional damping due to spin transfer into the NM.
This damping can be described by adding an additional damping torque ${\bm \tau}$ to the right-hand side of the LLG equation (\ref{eq:LLG}):\footnote{The torque in the direction of ${\bm m}\times (d{\bm m}/dt)$ is called the damping torque. In some junction systems, a torque in the direction of $d{\bm m}/dt$, which is called the field-like torque, becomes relevant.
Although we neglect the field-like torque for simplicity of discussion, we can relate it to the real part of the spin susceptibility, ${\rm Re} \, \chi^R({\rm 0},\omega_{\bm 0})$~\cite{Yama2023,Ominato2024} (see Sec.~\ref{sec:GilbertDamping}).}
\begin{align}
{\bm \tau} =-\delta \alpha_{\rm G} \, {\bm m} \times \frac{d{\bm m}}{dt}.
\label{eq_spin_mixing_conductance0}
\end{align}
Here, $\delta \alpha_{\rm G}$ ($>0$) is a modulation of the Gilbert damping.
The damping torque ${\bm \tau}$, which is equal to the loss of angular momentum per unit time per unit volume, is related to the spin current density across the junction, ${\bm j}_s$, as ${\bm \tau} = - {\bm j}_s/(\hbar \tilde{S}d)$, where $\tilde{S}$ is a spin per unit volume in the ferromagnet and $d$ is the thickness of the ferromagnet.

The spin current at the interface is usually evaluated using the following expression
\begin{align}
    \vb*{j}_s =
    \frac{\hbar}{4\pi}
    g^{\uparrow\downarrow}_r
    \vb*{m}\times\frac{d\vb*{m}}{dt}.
    \label{eq_spin_mixing_conductance}
\end{align}
Here, $g^{\uparrow\downarrow}_r$ is a parameter called the spin mixing conductance \cite{Tserkovnyak2002,Tserkovnyak2005}, with a dimension of the inverse of the area, which represents the efficiency of spin injection into the NM.
This method for spin current generation (or spin injection) driven by the continuous spin precession under microwave is called ``spin pumping''.
Combining Eqs.~(\ref{eq_spin_mixing_conductance0}) and (\ref{eq_spin_mixing_conductance}), we obtain $\delta \alpha_{\rm G} = g^{\uparrow \downarrow}_r/(4\pi \tilde{S} d)$.
This indicates that a thin ferromagnet is favorable for obtaining a large signal of spin transfer damping.

Although the spin-mixing conductance is a useful quantity for analyzing spin pumping, its use has several drawbacks.  
Since the spin-mixing conductance is theoretically formulated within scattering theory based on the adiabatic approximation~\cite{Tserkovnyak2002}, it cannot account for dynamical excitations in the target materials into which spins are injected.  
Furthermore, it is not well suited for obtaining microscopic insight.
These limitations of the conventional formulation based on spin-mixing conductance become significant when discussing temperature and frequency dependence, particularly when the characteristic frequencies of the target system are close to the FMR frequency.  
To address these issues, it is essential to adopt a formulation based on a microscopic tunneling Hamiltonian, which will be discussed in the next section.

\subsection{Microscopic description}
\label{sec:MicroscopicDescription}

In the following, let us consider a junction composed of a ferromagnetic insulator (FI) and a target material.
Although we choose an insulator for a ferromagnet here to avoid complications by a charge current path inside it, a similar calculation is possible also for a ferromagnetic metal.
Our microscopic Hamiltonian for spin transport at the magnetic interface is composed of three terms~\cite{ohnuma2014,ohnuma2017,Matsuo2018}
\begin{align}
    H =  \HFI + \Hsys + \Hex,
\end{align}
where $\HFI$, $H_{\mathrm{sys}}$, and  $\Hex$ describe the FI, the target material adjacent to the FI, and the interfacial exchange coupling, respectively.
In the following, we explain the details of $\HFI$ and $\Hex$, while the explicit forms of $H_{\mathrm{sys}}$ will be presented in Sec.~\ref{sec:superconductors}, Sec.~\ref{sec_SP_2DEG}, and Sec.~\ref{sec_SP_TMDC}.

The FI is modeled by the ferromagnetic Heisenberg model:
\begin{align}
    \HFI=-\mathcal{J}\sum_{\la n,m\ra}\vb*{S}_n\cdot\vb*{S}_m-\hbar\gamma_{\rm g} \hdc\sum_nS^z_n, \label{Heisenberg}
\end{align}
where $\mathcal{J}>0$ is the exchange coupling constant, $\la n,m\ra$ represents summation over all the nearest-neighbor sites, $\vS_n$ is the localized spin at site $n$ in the FI, and $\hdc$ is the static external magnetic field.
We define the Fourier transformation of the spin ladder operators, defined as $S^+_n = (N_{\rm FI})^{-1/2} \sum_{\bm k} S^+_{\bm k} e^{i{\bm k}\cdot {\bm r}_n}$ and $S^-_{\bm k}=(S^+_{\bm k})^\dagger$, where $N_{\rm FI}$ is a number of the localized spins in the FI, $S_n^+ = S_n^x + i S_n^y$, and ${\bm r}_n$ is a position of the site $n$.
Using the Holstein-Primakoff transformation \cite{holstein1940field}, these ladder operators are expressed with magnon annihilation and creation operators as 
\begin{align}
S_{\bm k}^+ = (2S_0)^{1/2} b_{\bm k}, \quad S_{\bm k}^- = (2S_0)^{1/2} b_{\bm k}^\dagger, \label{Skbk}
\end{align}
where $S_0$ is the amplitude of the spin per site.
Employing the spin-wave approximation, the Hamiltonian $\HFI$ is written as
\begin{align}
    \HFI\simeq\sum_{\vk}\hbar\omega_{\vk} b_{\vk}^\dagger b_{\vk},
    \label{HFImagnon}
\end{align}
where the constant terms are omitted. For a detail, see \ref{appA}.
We assume a parabolic dispersion $\hbar\omega_{\vk}=\mathcal{D}k^2+\hbar|\gamma_{\mathrm{g}}| h_{\mathrm{dc}}$ with a spin stiffness constant $\mathcal{D}$.
We consider spin dynamics under microwave irradiation. The interaction between the localized spin and the microwave is given by
\begin{align}
V(t)=\frac{\hbar\gamma_{\rm g}\hac}{2} \sum_n\qty(S^+_n e^{i\omega t} + S^-_n e^{-i\omega t}),
\end{align}
where $\hac$ and $\omega$ are respectively the amplitude and frequency of the microwave.
For the isolated FI ($H_{\rm ex}=0$), the amplitude of the small spin precession induced by microwave is calculated by the linear response theory as~\cite{Funato2022}
\begin{align}
\langle S^+_{\rm tot} \rangle = \frac{\hbar \gamma_{\rm g} N_{\rm FI}}{2} G^R({\bm 0},\omega) h_{\rm ac} e^{-i\omega t},
\end{align}
where $S^{\pm}_{\rm tot} = \sum_n S_n^\pm$.
The retarded spin correlation function $G^R({\bm k},\omega)$ is defined as
\begin{align}
G^R({\bm k},\omega) &= \int dt \, e^{i\omega t} G^R({\bm k},t) ,\\
G^R({\bm k},t) &= -i\theta(t) \langle [S_{\bm k}^+(t), S_{\bm k}^-(0)]\rangle,
\end{align}
where $\theta(t)$ is the Heaviside step function.
Using Eq.~(\ref{Skbk}), the spin correlation function $G^R({\bm k},\omega)$ can be regarded as the magnon propagator.
It is calculated from the Hamiltonian (\ref{HFImagnon}) as
\begin{align}
G^R({\bm k},\omega) = \frac{S_0/\hbar}{\omega-\omega_{\bm k}+i\delta}.
\end{align}
For real FIs, the Gilbert damping is in general caused by magnon-magnon and magnon-phonon interactions at finite temperatures.
Since the microscopic derivation of the Gilbert damping is very complicated, we simply replace an infinitesimal $\delta$ with $\alpha_{\rm G}\omega$ in the following calculation, where $\alpha_{\rm G}$ is a phenomenological parameter~\cite{Kasuya1961,Cherepanov1993,Jin2019}.
Then, the spin correlation function is related to the linear response coefficient given in Eq.~(\ref{chiLLG}) as $\chi_{\rm FI} =  (\hbar|\gamma_{\rm g}| N_{\rm FI}/2) G^R({\bm 0},\omega)$.
This relation holds even in the presence of the interfacial coupling explained in the following.

The proximity exchange coupling at the interface, $\Hex$, is given by  
\begin{align}
    &\Hex = \HZ + \HT, \label{Hex1} \\
    &\HZ = \sum_{\vk,\vq} J_{\vk,\vq} S^z_{\vk} s^z_{\vq}, \label{Hex2}\\
    &\HT = \frac{1}{2} 
    \sum_{\vk,\vq} \left( J_{\vk,\vq} S^+_{\vk} s^-_{\vq} + \mathrm{h.c.} \right). \label{eq_Hex}
\end{align}
where $\HZ$ describes the effective Zeeman field along the precession axis, and $\HT$ represents spin transfer between the FI and the system.
Explicit forms of the spin operators in the target material, $s^z_{\bm q}$ and $s^\pm_{\bm q}$, will be shown later.
While we have assumed that the precession axis is in the $z$ direction here, its direction will be changed depending on the system in the subsequent sections.
Since the matrix element $J_{\vk,\vq}$ generally depends on interface roughness, we have to perform averaging with respect to randomness or surface roughness after calculating physical quantities.
By a simple treatment of the interface~\cite{Ominato2022,Ominato2022b,Yama2025}, however, clean and dirty interfaces can be modeled by setting $J_{\vk,\vq}$ as~\footnote{Note that $J$ includes not only the strength of the interfacial exchange coupling but also the number of the unit cells in the FI and the target material. For a detail, see Refs.~\cite{Ominato2022,Ominato2022b,Yama2025}.}
\begin{align}
    &\text{Clean interface:} \quad J_{\vk,\vq} = J \delta_{\vk,\vq}, \label{eq_averaged_J2} \\
    &\text{Dirty interface:} \quad J_{\vk,\vq} = J. \label{eq_averaged_J1}
\end{align}
The first case corresponds to a flat interface, where momentum is conserved during the interaction.
The second case corresponds to a rough interface, where momentum is not conserved and transitions to all momenta are allowed, assuming that the matrix elements are equal for all processes.
We note that the modeling of the interface is one of the important issues in discussing spin transport in magnetic junctions.
Realistic modeling has been discussed in a few recent theoretical studies~\cite{Yama2025,Heydari2025}.

\subsection{Enhanced Gilbert damping}
\label{sec:GilbertDamping}

Since the retarded spin correlation function $G^R({\bm k},\omega)$ corresponds to the magnon propagator, we can use a standard technique of the Green's function method.
In the presence of the interfacial exchange coupling, $G^R({\bm k},\omega)$ is generally expressed from the Dyson equation as
\begin{align}
    G^R(\vk,\omega)=\frac{2S_0/\hbar}{\omega-\omega_{\vk}+i\alpha_{\mathrm{G}}\omega-(2S_0/\hbar)\Sigma^R(\vk,\omega)}.
\end{align}
The self-energy $\Sigma^R(\vk,\omega)$ due to the interfacial coupling is calculated by second-order perturbation as
\begin{align}
    \Sigma^R(\vk,\omega)=
    \sum_{\vq}
    |J_{\vk,\vq}|^2
    \chi^{R}(\vq,\omega),
\end{align}
where $\chi^{R}(\vq,\omega)$ is the dynamic spin susceptibility of the target material defined as
\begin{align}
\chi^R(\vq,\omega)=\int dt \, e^{i\omega t}\chi^R(\vq,t), \\
\chi^R(\vq,t)=-\frac{i}{\hbar} \theta(t) \langle [ s^+_{\vq}(t), s^-_{\vq}(0) ] \rangle .
\end{align}
By rewriting the spin correlation function as
\begin{align}
G^R({\bm 0},\omega)=\frac{2S_0/\hbar}{\omega-\omega_{\bm 0}+i(\alpha_{\mathrm{G}}+\delta \alpha_{\rm G})\omega},
\end{align}
and by using the modeling of the interface in Sec.~\ref{sec:MicroscopicDescription}, the enhanced Gilbert damping $\delta\alpha_{\mathrm{G}}$ is calculated for a clean interface as
\begin{align}
    \delta\alpha_{\mathrm{G}}
    &\simeq -\frac{2S_0}{\hbar\omega_{\bm 0}}|J|^2\, {\rm Im}\, \chi^R({\bm 0},\omega_{\bm 0}),
\end{align}
and for a dirty interface as
\begin{align}
    \delta\alpha_{\mathrm{G}}
    &\simeq -\frac{2S_0}{\hbar\omega_{\bm 0}}|J|^2 \, {\rm Im}\, \sum_{\bm q} \chi^R({\bm q},\omega_{\bm 0}).
\end{align}
Here we have assumed that a FMR peak is sharp enough ($\alpha_{\rm G}+\delta \alpha_{\rm G}\ll 1$).
We emphasize that the bulk Gilbert damping, $\alpha_{\rm G}$, is treated phenomenologically, whereas the modulation of the bulk Gilbert damping, $\delta \alpha_{\rm G}$, can be calculated from our microscopic model.
The present microscopic calculation reveals the physical interpretation of the enhancement of the Gilbert damping: $\delta \alpha_{\rm G}$ is related to the imaginary part of the uniform (local) spin susceptibility of the target material for a clean (dirty) interface.~\footnote{We note that the real part of the self-energy has been absorbed into the FMR frequency $\omega_{\bm 0}$ for simplicity. However, the shift of the FMR frequency also includes information on the target spin susceptibility and is interpreted in part as the effect of the field-like torque in the context of the spin mixing conductance. For a detailed discussion, see, e.g, Refs.~\cite{Ominato2022b,Yama2023}.}
Our interface model and the corresponding expressions of $J_{\vk,\vq}$ and $\delta\alpha_{\mathrm{G}}$ are summarized in Table~\ref{table:chi_uni}.

\begin{table}[b]
\centering
\begin{tabular}{l||c|c}
    \hline
    Interface &
    Clean &
    Dirty \\
    \hline\hline
    $J_{\vk,\vq}$ &
    $J\delta_{\vk,\vq}$ &
    $J$ \\
    \hline
    $\delta \alpha_{\mathrm{G}}$ &
    $\mathrm{Im}\chi^R(\vb*{0},\omega)$ &
    $\mathrm{Im}\sum_{\vq}\chi^R(\vq,\omega)$ \\
    \hline
\end{tabular}
\caption{
Expressions of $J_{\vk,\vq}$ and $\delta \alpha_{\mathrm{G}}$ for either clean or dirty interface.
}
\label{table:chi_uni}
\end{table}

As an illustrative example, let us consider a normal metal (NM) as the target material using a model of non-interacting electrons:
\begin{align}
H_{\rm sys} &= \sum_{{\bm k}\sigma} \xi_{\bm k} c_{{\bm k}\sigma}^\dagger 
c_{{\bm k}\sigma},
\end{align}
where $\xi_{\bm k}$ is an electron energy measured from a chemical potential.
The spin operators in Eqs.~(\ref{Hex1})-(\ref{eq_Hex}) are given as
\begin{align}
s_{\bm q}^{a} &= \frac12 \sum_{\sigma\sigma'}
\sum_{{\bm k}} c_{{\bm k}\sigma}^\dagger (\hat{\sigma}_a)_{\sigma \sigma'}
c_{{\bm k}+{\bm q}\sigma'}, \quad (a=x,y,z),\label{sqdef} \\
s_{\bm q}^+ &= s_{\bm q}^x + i s_{\bm q}^y = \sum_{\bm k} c_{{\bm k}\uparrow}^\dagger c_{{\bm k}+{\bm q}\downarrow} , 
\end{align}
and $s_{\bm q}^{-} = (s_{\bm q}^+)^\dagger$, where $\hat{\sigma}_a$ ($a=x,y,z$) are the Pauli matrices.
The spin susceptibility is calculated as
\begin{align}
\chi^R({\bm q},\omega) = \sum_{{\bm k}} \frac{f(\xi_{{\bm k}+{\bm q}})-f(\xi_{{\bm k}})}{\hbar \omega +\xi_{\bm k} - \xi_{{\bm k}+{\bm q}}+i\delta}.
\end{align}
The enhancement of the Gilbert damping vanishes at a finite $\omega_{{\bm 0}}$ for a clean interface, while it is calculated for a dirty interface as
\begin{align}
\delta \alpha_{\rm G} = 2\pi S_0 |J|^2 D_{\rm n}(\epsilon_{\rm F})^2,
\end{align}
where $D_{\rm n}(\epsilon_{\rm F})$ is a density of states per unit cell in the NM.

\subsection{Spin current generation}

As discussed in Sec.~\ref{sec:SPSMC}, the increase of the Gilbert damping indicates the spin current generation across the junction.
This spin current is calculated by the method of the nonequilibrium Green's function in the Keldysh formalism as follows.
Let us consider a normal metal described by noninteracting electrons.
We first define the total spin as
\begin{align}
    s^z_{\mathrm{tot}}:= \frac12 \sum_{\vk} (c_{\vk \uparrow}^\dagger c_{\vk \uparrow} - c_{\vk \downarrow}^\dagger c_{\vk \downarrow}) .
\end{align}
Then, the spin current operator is defined as
\begin{align}
    \IS &=i\qty[s^z_{\mathrm{tot}},H] \nonumber \\
&=\frac{i}{4}\sum_{\vq}\qty[J_{\vk,\vq}S^-_{\vk}s^+_{\vq}-\mathrm{h.c.}].
\end{align}
We calculate the statistical average of the spin current at the interface and treat $\HT$ as a perturbation and $\HFI+\Hsys+\HZ$ as an unperturbed Hamiltonian. The second-order perturbation calculation with respect to $\HT$ gives the statistical average of the spin current at the interface~\cite{ohnuma2014,Bruus2004,ohnuma2017,Matsuo2018,Kato2019}
\begin{align}
    \la \IS(t) \ra=\mathrm{Re}\qty[\frac{i}{2}\sum_{\vk,\vq}J_{\vk,\vq}\la S_{\vk}^-(t)s_{\vq}^+(t)\ra].
\end{align}
Within the second-order perturbation with respect to the interaction between the localized spin and the microwave $V(t)$, the spin current is given by
\begin{align}
    \la \IS\ra \simeq
    \frac{|J|^2{\cal A} (S_0\gamma_{\rm g}\hac)^2}{(\omega-\omega_{\bm 0})^2+\alpha^2_{\mathrm{G}}\omega^2}\, \mathrm{Im}\,\chi^{R}(\vb*{0},\omega),
    \label{eq_tunneling_spin_current}
\end{align}
for a clean interface, while $\chi^{R}(\vb*{0},\omega)$ is replaced with $\sum_{{\bm q}} \chi^{R}(\vb*{q},\omega)$ for a dirty interface.
For a detail of the calculation, see \ref{appB}.
Here, ${\cal A}$ is the interface area and we have assumed that the FMR peak is sharp enough ($\alpha_{\mathrm{G}}+\delta\alpha_{\mathrm{G}} \ll 1$).
From this expression, one can see that the interface spin current is characterized by the dynamic spin susceptibility of the adjacent paramagnetic material~\footnote{
We note that the nonequilibrium fluctuation of the spin current induced by the spin pumping includes information of a unit of the angular momentum carried by magnetic excitations~\cite{Kamra2014,Kamra2016,Kamra2016b,Sato2025}.}.

The spin current induced by spin pumping can be detected using the inverse spin Hall effect (ISHE) \cite{Saitoh2006,Valenzuela2006,Zhao2006,Kimura2007,Ando2011}, by which the spin current is converted into a charge current as ${\bm j}_{\rm c} = \theta_{\rm SH} (e/\hbar) {\bm j}_s \times \hat{\bm n}$ via the spin-orbit interaction in the target material, where $\theta_{\rm SH}$ is the spin Hall angle and $\hat{\bm n}$ is a unit normal vector of the interface.
To treat such spin-charge conversion, we need to combine another theoretical method for describing electric states in the target material.
In Sec.~\ref{sec_SP_2DEG}, we treat the inverse Rashba–Edelstein effect (IREE), using the Boltzmann equation.

\subsection{Experimental relevance}

This subsection briefly summarizes the experimental reports on spin pumping into two-dimensional systems. As mentioned in the Introduction, spin pumping is a useful experimental tool for obtaining information on spin excitations in two-dimensional systems, because of its surface sensitivity. Most of the experimental analyses reported so far have relied on the spin-mixing conductance, which does not directly reveal the connection to the underlying microscopic electronic states, while useful for quantifying spin transfer across interfaces. Keeping this in mind, we first introduce several experiments on graphene and outline their development. We then describe recent progress in experiments on TMDC materials, and finally comment on reports involving the surface states of three-dimensional topological insulators.

\begin{figure}
\begin{center}
\includegraphics[width=0.8\hsize]{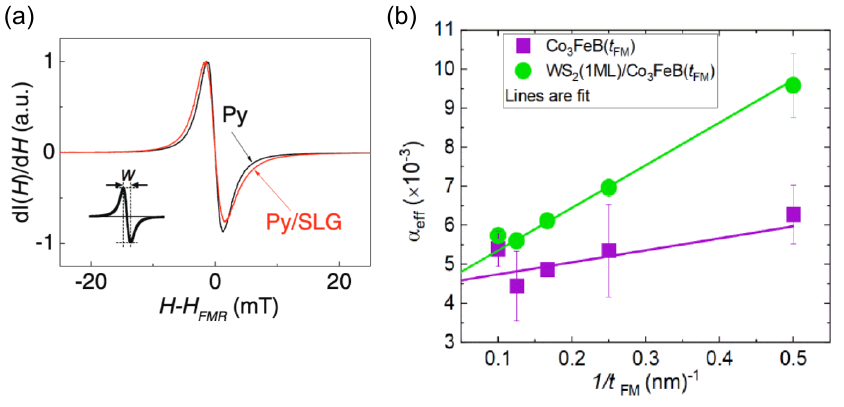}
\end{center}
\caption{(a) Broadening of the FMR linewidth due to spin transfer at the interface between 25-nm-thick permalloy (Py) and single-layer graphene. Measurements were performed at room temperature and microwave frequency $9.62~\mathrm{GHz}$.
Adapted from Ref.~\cite{Tang2013}. (b) Dependence of the effective damping enhancement on the Co$_3$FeB thickness ($t_{\mathrm{FM}}$) in a WS$_2$/Co$_3$FeB heterostructure. Adapted from Ref.~\cite{Husain2022}.}
\label{fig_FMR_signal_2d_system}
\end{figure}

The FMR damping enhancement induced by spin pumping has been reported for permalloy (Py)/graphene junction systems~\cite{Patra2012,Tang2013,Singh2013}. Figure~\ref{fig_FMR_signal_2d_system} (a) shows an example of the FMR experiment for Py/graphene~\cite{Tang2013}.
The horizontal axis is an external magnetic field, while the vertical axis indicates the derivative of the microwave absorption with respect to the magnetic field.
We note that the origin of the horizontal axis is chosen as the magnetic field for the FMR resonance, $H_{\rm FMR}$.
We find that the peak of the microwave absorption becomes broader for a Py/graphene junction compared with an isolated Py. In addition to the FMR damping enhancement, the same group also observed an inverse spin Hall voltage generated by the injected spin current and reported this as evidence for spin injection into graphene. However, it should be noted that when discussing damping enhancement attributed to spin pumping, structural modifications of the ferromagnet induced by forming the heterostructure can also give the appearance of increased damping~\cite{Berger2014}, and the experimental results should therefore be interpreted with particular caution. Since these pioneering spin pumping experiments on graphene, various experimental groups have reported spin pumping related phenomena, and it is becoming a consensus that spin pumping into graphene is possible and can be characterized by FMR modulation~\cite{Ohshima2014,Mendes2015,Dushenko2016,Indolese2018,Mendes2019,Park2020,Cunha2024}.

Similar experimental observations have also been reported by several experimental groups in TMDC materials~\cite{husain2018,mendes2018,bansal2019,Wu2020,Husain2022,Bangar2022,Victor2025}. Figure~\ref{fig_FMR_signal_2d_system}~(b) shows the effective damping constant of a WS$_2$/Co$_3$FeB (FM) heterojunction plotted as a function of the inverse FM thickness $1/t_{\mathrm{FM}}$. From this result, the authors argued that the FM exhibits a damping enhancement when interfaced with WS$_2$, and that the proportionality to $1/t_{\mathrm{FM}}$ is consistent with spin pumping in the heterojunction. On the other hand, a very recent study reports that in an atomically flat MoSe$_2$/CoFeB heterojunction, no damping enhancement attributable to spin pumping was detected~\cite{Lu2025}. Careful analysis of the experiments, together with a microscopic understanding of the interfacial electronic states, will be essential for achieving a consistent interpretation of these findings.

Spin transport via two-dimensional Dirac fermions on the surfaces of three dimensional topological insulators has been explored using ferromagnet and topological insulator heterostructures~\cite{Shiomi2014,Wang2016,Baker2015,Mendes2017,Tang2018}. In these systems, the spin momentum locked surface states enable efficient spin to charge conversion and modify the magnetization dynamics of the adjacent magnetic layer. Spin pumping into these surface states has been reported to cause a noticeable enhancement of the effective damping, and signatures of inverse Edelstein effect and spin torque generation have also been observed. These results indicate that topological insulator surfaces offer a promising platform for spin current generation and detection.

Experimental studies on graphene, TMDCs, and three-dimensional topological insulators have advanced considerably, and the description in terms of spin mixing conductance has played an important role in quantifying spin transfer in the early stages. This approach, however, does not directly reveal the connection to characteristic microscopic electronic structures such as massless Dirac fermions, spin valley coupling, or spin momentum locking. To gain deeper insight, it is important to develop and apply microscopic theoretical descriptions alongside experimental progress so that both can advance together. The following sections provide such a theoretical framework for spin pumping into specific two-dimensional systems.

\section{Spin pumping into superconductors}
\label{sec:superconductors}

In superconductivity research, determining the pairing symmetry of Cooper pairs is a fundamental issue.
In particular, probing the spin excitations of the superconducting state is essentially important, because the nature of the spin susceptibility reflects the characteristics of the Cooper pairs.
As explained in the previous section, when FMR occurs in a ferromagnetic thin film stacked with a metallic layer, spin excitations are induced in the conduction electrons of the metal layer.
As a result, the magnetization dynamics is modulated, and the FMR signal obtained from the bilayer reflects this modulation, which provides information about the spin susceptibility.
Conventional spin probes, such as NMR and neutron scattering, have been widely used for this purpose.
However, these methods suffer from insufficient sensitivity when applied to thin films.
FMR modulation, which is highly sensitive to interfaces and thin-film structures, offers a promising alternative for investigating two-dimensional superconductors \cite{Ye2012,Xi2016,Cao2018,Park2021,Guo2004,Wang2012,Weerdenburg2023,Saito2017}.
These techniques are complementary, allowing for a more comprehensive investigation of superconducting properties.
NMR probes the MHz regime, while FMR operates in the GHz regime, enabling the study of spin responses at higher frequencies than conventional techniques.

\subsection{Spin-singlet superconductors}

\begin{figure}
\begin{center}
\includegraphics[width=0.7\hsize]{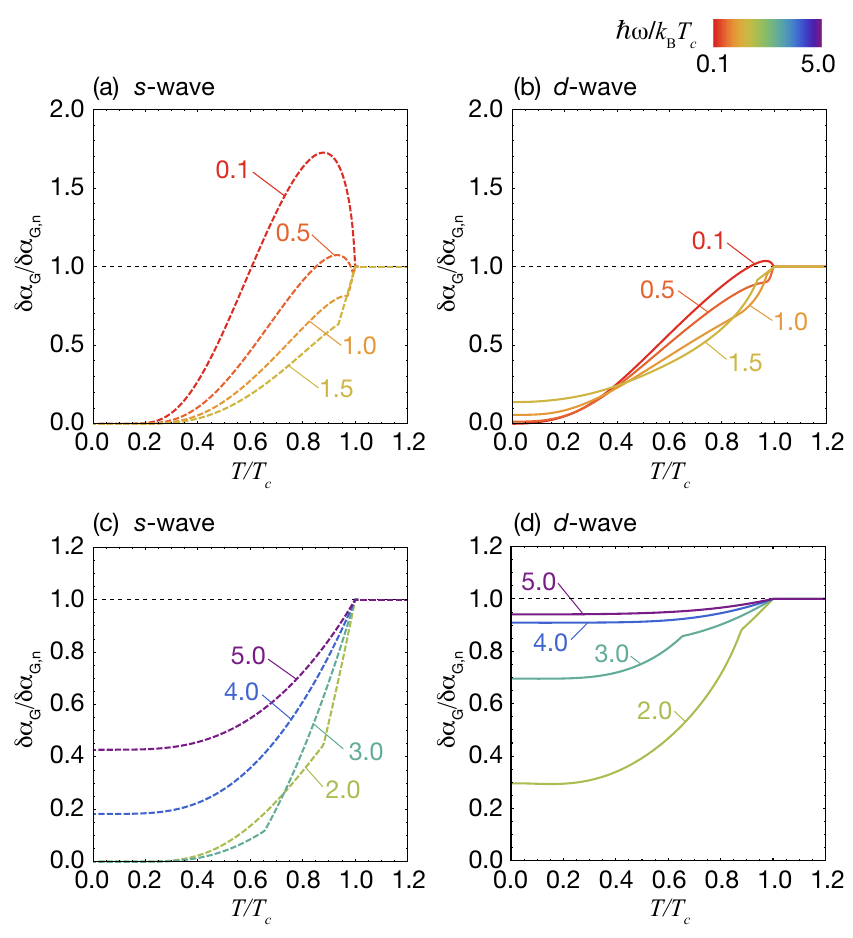}
\end{center}
\caption{Enhanced Gilbert damping $\delta\alpha_{\mathrm{G}}$ as a function of temperature $T$.  
(a) and (c) show $\delta\alpha_{\mathrm{G}}$ in the $s$-wave SC in the low- and high-frequency cases, respectively.  
(b) and (d) show $\delta\alpha_{\mathrm{G}}$ in the $d$-wave SC in the low- and high-frequency cases, respectively.  
$\delta\alpha_{\mathrm{G},n}$ is the normal-state value.
Adapted from Ref\cite{Ominato2022}.
}
\label{fig_FMR_dSC}
\end{figure}

Let us first consider spin pumping into two-dimensional spin-singlet superconducting states \cite{inoueSpinPumpingSuperconductors2017,Kato2019,Silaev2020,Ominato2022}, with a focus on comparing $s$-wave and $d$-wave pairing.
The system Hamiltonian is given as
\begin{align}
H_{\rm sys} = \sum_{{\bm k}} ( c^\dagger_{{\bm k}\uparrow} \ c_{-{\bm k}\downarrow} )
\left( \begin{array}{cc}
\xi_{\bm k} & \Delta_{\bm k} \\ \Delta_{\bm k} & -\xi_{\bm k}
\end{array} \right) 
\left( \begin{array}{c} 
c_{{\bm k}\uparrow} \\ c^\dagger_{-{\bm k}\downarrow}
\end{array} \right) .
\end{align}
The pair potentials are given respectively by $\Delta_{\vb*{k}} = \Delta$ for the $s$-wave and $\Delta_{\vb*{k}} = \Delta \cos 2\phi_{\vb*{k}}$ for the $d$-wave states.
We employ a phenomenological expression $\Delta = 1.76\,k_{\mathrm{B}} T_c \tanh(1.74\sqrt{T_c/T - 1})$ in the present calculation. 
In spin-singlet superconductors, spin is a conserved quantity, as in the normal state.
Consequently, the uniform spin susceptibility $\chi^R(\vb*{0},\omega)$ vanishes, and the Gilbert damping modulation $\delta\alpha_{\mathrm{G}}$ is governed by the local spin susceptibility $\sum_{\vb*{q}} \chi^R(\vb*{q},\omega)$.
By a straightforward calculation, we obtain the expression for $\delta\alpha_{\mathrm{G}}$ as
\begin{align}
    \delta\alpha_{\mathrm{G}} =
    \frac{2\pi S_0J^2 D_{\mathrm{n}}(\epsilon_{\rm F})^2}{\hbar\omega}
    \int_{-\infty}^{\infty} dE\,
    [f(E) - f(E+\hbar\omega)]
    F(E,\omega)
    D(E) D(E+\hbar\omega),
\end{align}
where $F(E,\omega)$ is the coherence factor given by
\begin{align}
    F(E,\omega) =
    \begin{cases}
        1 + \dfrac{\Delta^2}{E(E+\hbar\omega)} & \text{: $s$-wave} \\
        1 & \text{: $d$-wave}
    \end{cases}
\end{align}
and $D(E)$ is the density of states of Bogoliubov quasiparticles,
\begin{align}
    D(E) =
    \begin{cases}
        \mathrm{Re}\left(\dfrac{|E|}{\sqrt{E^2 - \Delta^2}}\right) & \text{: $s$-wave} \\
        \mathrm{Re}\left[\dfrac{2}{\pi} K\left(\dfrac{\Delta^2}{E^2}\right)\right] & \text{: $d$-wave}
    \end{cases}
\end{align}
with $K(x)$ being the complete elliptic integral of the first kind,
\begin{align}
    K(x) = \int_0^{\pi/2} \frac{d\phi}{\sqrt{1 - x \cos^2\phi}}.
\end{align}
These expressions show that $\delta\alpha_{\mathrm{G}}$ directly reflects the density of states.
Figure~\ref{fig_FMR_dSC} shows $\delta\alpha_{\mathrm{G}}$ as a function of $T$.
For both $s$-wave and $d$-wave pairings, a coherence peak appears just below the superconducting transition temperature when the microwave frequency is small compared to $k_{\mathrm{B}} T_c$ (i.e., $\hbar\omega \ll k_{{\mathrm{B}}} T_c$), though the peak is less pronounced in the $d$-wave superconductor.
Near zero temperature, $\delta\alpha_{\mathrm{G}}$ approaches zero in both cases. $\delta\alpha_{\mathrm{G}}$ decays exponentially for the $s$-wave superconductor, while $\delta\alpha_{\mathrm{G}}$ follows a power law $\delta\alpha_{\mathrm G} \propto T^2$ for the $d$-wave superconductor.
This difference arises from the distinct gap structures of the Bogoliubov quasiparticles, which means the full gap for $s$-wave and the nodal structure for $d$-wave superconductors.
As the frequency increases, the coherence peak is suppressed. Near zero temperature, $\delta\alpha_{\mathrm{G}}$ remains finite in the $d$-wave superconductor due to the residual density of states associated with nodal quasiparticles, whereas $\delta\alpha_{\mathrm{G}}$ is exponentially small for the $s$-wave superconductor.
This behavior again reflects the presence or absence of a gap in the quasiparticle spectrum.
At sufficiently high frequencies, even below $T_c$, the $\delta\alpha_{\mathrm{G}}$ approaches its normal-state value.
These distinct temperature and frequency dependencies in $s$-wave and $d$-wave superconductors indicate that FMR modulation measurements can serve as a useful probe for identifying pairing symmetry in nanoscale superconductors such as atomically thin materials.

\subsection{Spin-triplet superconductors}

\begin{figure}
\begin{center}
\includegraphics[width=0.7\hsize]{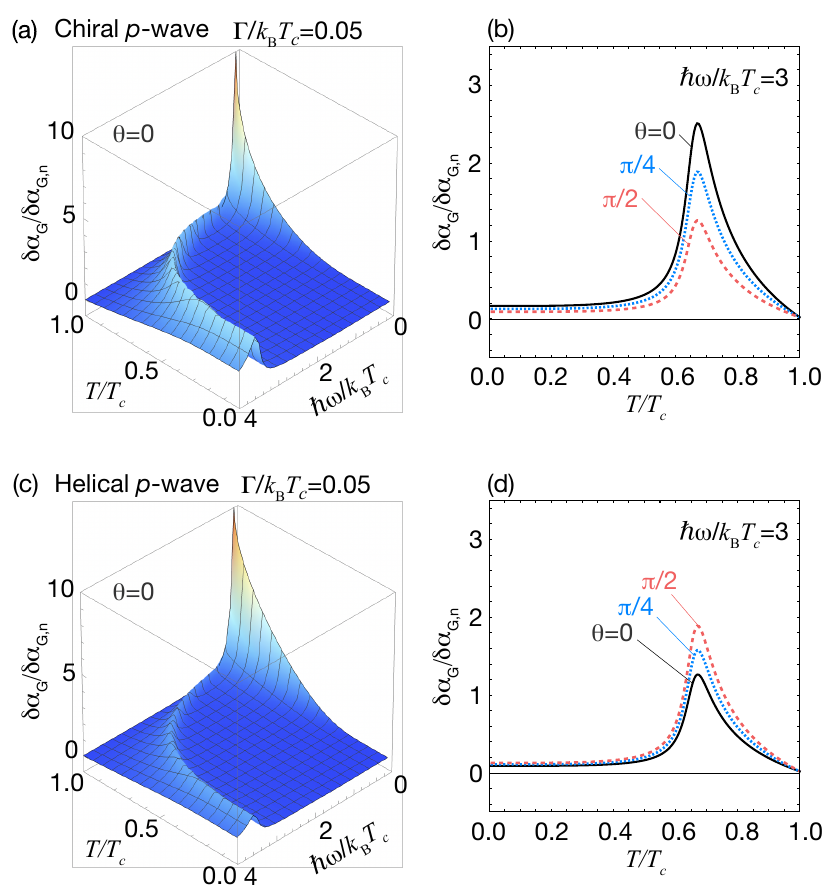}
\end{center}
\caption{The enhanced Gilbert damping for (a),(b) Chiral $p$-wave SC and (c),(d) Helical $p$-wave SC.
$\delta\alpha_{\mathrm{G},n}=S_0J^2D_{\mathrm{n}}(\epsilon_{\mathrm{F}})/(N_{\mathrm{FI}}k_{\mathrm{B}}T_c)$ is the characteristic value in the normal state.
Adapted from Ref \cite{Ominato2022b}.
}
\label{fig_FMR_pSC}
\end{figure}

Next, we turn to two-dimensional spin-triplet $p$-wave superconducting states \cite{Ominato2022b}.
Here, we consider two representative cases: the chiral pairing, where the Cooper pair spin lies in the plane, and the helical pairing, where it points out of plane. These are rotationally symmetric $p$-wave states with an energy spectrum identical to that of the $s$-wave superconductor, thus serving as minimal models for identifying features unique to spin-triplet superconductors.
The pair potential is expressed as $\Delta_{\vb*{k}} = \vb*{d} \cdot \vb*{\sigma} i\sigma_y$, where the vector $\vb*{d}$ characterizes the spin orientation of the Cooper pairs.
The explicit forms of $\vb*{d}$ differ for the chiral and helical pairing states,
\begin{align}
    \vb*{d}=
    \begin{cases}
        \Delta(0,0,e^{i\phi_{\vk}}) & :\mathrm{Chiral} \\
        \Delta(-\sin\phi_{\vk},\cos\phi_{\vk},0) & :\mathrm{Helical}
    \end{cases}
\end{align}
In the following analysis, we consider a flat interface model, and $\delta\alpha_{\mathrm{G}}$ is determined by the uniform spin susceptibility $\chi^R(\vb*{0},\omega)$.
In contrast to spin-singlet superconductors, spin is not a conserved quantity in spin-triplet superconductors, allowing $\chi^R(\vb*{0},\omega)$ to remain finite.
Figure~\ref{fig_FMR_pSC} shows $\delta\alpha_{\mathrm{G}}$ for both chiral and helical pairing states. 
To avoid divergence under a resonance condition, we introduce a finite quasiparticle lifetime $\Gamma$.
We find that $\delta\alpha_{\mathrm{G}}$ exhibits a resonance peak under the resonance condition $\hbar\omega = 2\Delta$.
This peak is a distinct feature of spin-triplet states that is absent in spin-singlet states.
The distinction between chiral and helical states appears in the angular dependence of $\delta\alpha_{\mathrm{G}}$ with respect to the magnetization direction.
Defining $\theta$ as the angle between the magnetization and the interface normal, we present the temperature dependence of $\delta\alpha_{\mathrm{G}}$ for several values of $\theta$ in Figs.~\ref{fig_FMR_pSC}(b) and \ref{fig_FMR_pSC}(d).
The results show that in the chiral superconductor, $\delta\alpha_{\mathrm{G}}$ decreases with increasing $\theta$, whereas in the helical superconductor, it increases with $\theta$.
This contrast allows the spin orientation of the triplet Cooper pairs to be identified through the angular dependence of $\delta\alpha_{\mathrm{G}}$.

\subsection{Recent development}

\begin{figure}
\begin{center}
\includegraphics[width=0.6\hsize]{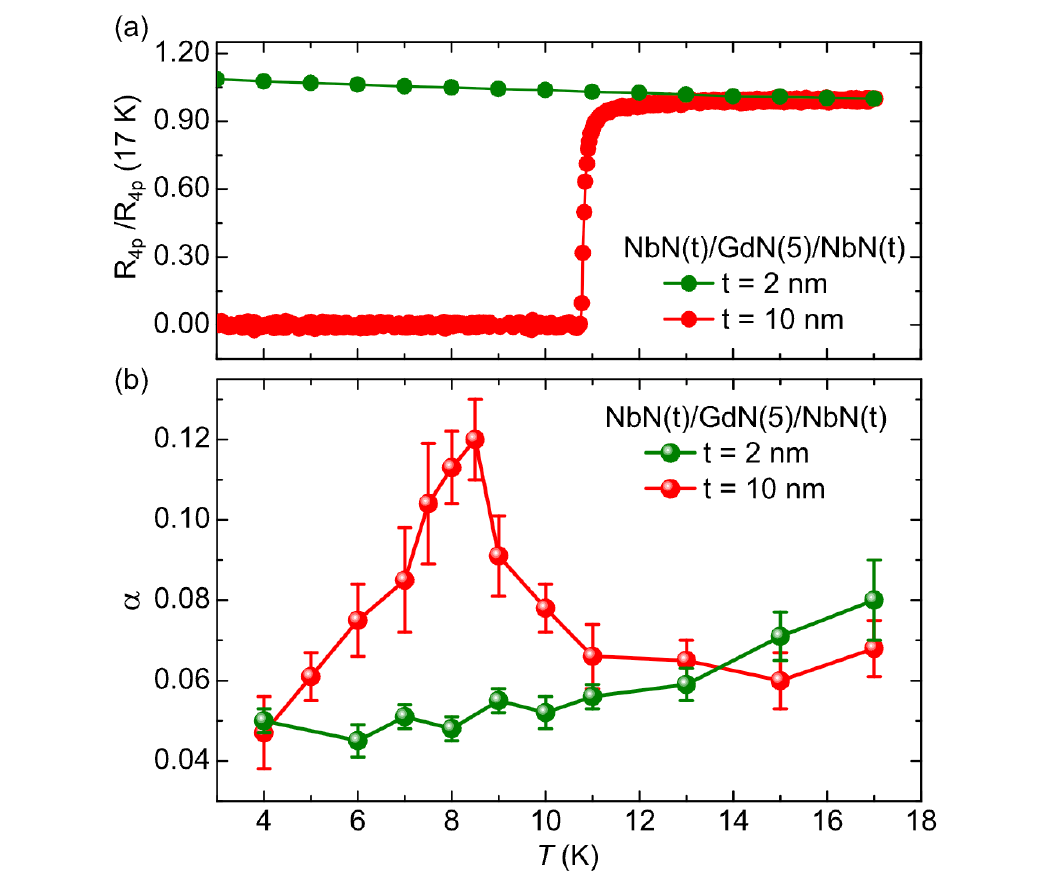}
\end{center}
\caption{Spin dynamics of the superconducting NbN thin films probed via spin pumping. The normalized four-probe resistance (a) and Gilbert damping (b) as a function of the temperature for the samples of NbN (2)/GdN (5)/NbN (2) and NbN (10)/GdN (5)/NbN (10), respectively.
Adapted from Ref \cite{Yao2018}.
}
\label{fig_FMR_damping_modulation_SC}
\end{figure}

Finally, we briefly summarize recent theoretical and experimental developments relevant to spin pumping into superconductors.
On the theoretical side, related studies have addressed a range of topics, including spin pumping into ferromagnetic superconductors \cite{Funaki2023}, Majorana Ising spin dynamics excited by FMR~\cite{Ominato2024}, spin relaxation in $s$-wave superconductors~\cite{Silaev2020}, the influence of Andreev bound states on spin pumping at superconductor/FI interfaces~\cite{Silaev2020b,Sun2023b}, spin pumping via antiferromagnetic resonance in $s$-wave superconductor/antiferromagnetic insulator junctions~\cite{Holm2021}, magnon current generation driven by spin-triplet spin currents~\cite{Johnsen2021}, the effect of superconducting fluctuations on the spin Hall effect~\cite{Watanabe2022}, spin dynamics in superconductor/FI hybrid structures~\cite{Turkin2023,Ojajarvi2022}, and magnon-cooparons in magnet-superconductor hybrids~\cite{Bobkova2022}. Theoretical proposals also include controlling FMR via superconducting environments~\cite{Zhou2023}.
On the experimental side, enhanced FMR linewidths have been reported in magnetic multilayers involving $s$-wave superconductors and FIs~\cite{Yao2018}, as well as in $d$-wave superconductor/ferromagnetic metal junctions~\cite{Carreira2020}. Recent experiments demonstrated spin pumping into $s$-wave superconductors mediated by triplet Cooper pairs~\cite{Vanstone2023}, FMR shifts induced by superconducting proximity effects~\cite{Li2025}, and magnetic coupling across superconducting spacers mediated by Yu--Shiba--Rusinov bound states~\cite{Lu2024}.
A representative experiment observing a coherence peak in the FMR linewidth of an $s$-wave superconductor is shown in Fig.~\ref{fig_FMR_damping_modulation_SC}.
Our results suggest that FMR modulation measurements can play a significant role in elucidating superconducting pairing symmetry and contribute to the development of spectroscopic techniques for nanoscale magnetic heterostructures.

\section{Spin pumping into 2DEG}
\label{sec_SP_2DEG}

Semiconductor technology has been an attractive choice for realizing integrated spintronic devices~\cite{Fabian2007,Kohda2017}. 
Two-dimensional electron gas (2DEG) fabricated in semiconductor heterostructures has provided an ideal stage for spintronic devices.
The field-effect spin transistor, proposed by Datta and Das~\cite{Datta1990}, is a famous example.
In such devices, two kinds of spin-orbit interactions play an important role.
One is the Rashba spin-orbit interaction, which appears in systems with broken structural inversion symmetry, mainly induced by confinement potentials in the stacking direction~\cite{Bychkov1984, Rashba2015}.
The other is the Dresselhaus spin-orbit interaction~\cite{Dresselhaus1955}, which arises when the crystal structure lacks inversion symmetry, as in zinc-blende III-V and II-VI semiconductors.
Both spin-orbit interactions induce spin-momentum locking, i.e., strong dependence of the effective Zeeman field on the electron propagation direction.
It is notable that both Rashba and Dresselhaus interactions can coexist and their relative strength can be tuned by a gate voltage~\cite{Nitta1997,Kohda2017,Manchon2015}.
When the strengths of the Rashba and Dresselhaus SOCs are equal, a persistent spin helix (PSH) state emerges~\cite{Bernevig2006, Weber2007, Koralek2009, Kohda2012, Sasaki2014, Nitta2023}, leading to prolonged electron spin lifetimes. 
The Rashba spin-orbit interaction becomes significant also at an interface in the junction systems or heterostructures of transition oxides~\cite{Chen2023}.

In the viewpoint of application, it is an attractive strategy to combine spin pumping with spin-dependent transport in 2DEG. 
In the following, we review recent theoretical studies on spin pumping into 2DEG with both Rashba and Dresselhaus spin-orbit interactions.

\subsection{Model}

\begin{figure}[tb]
\centering
\includegraphics[width=140mm]{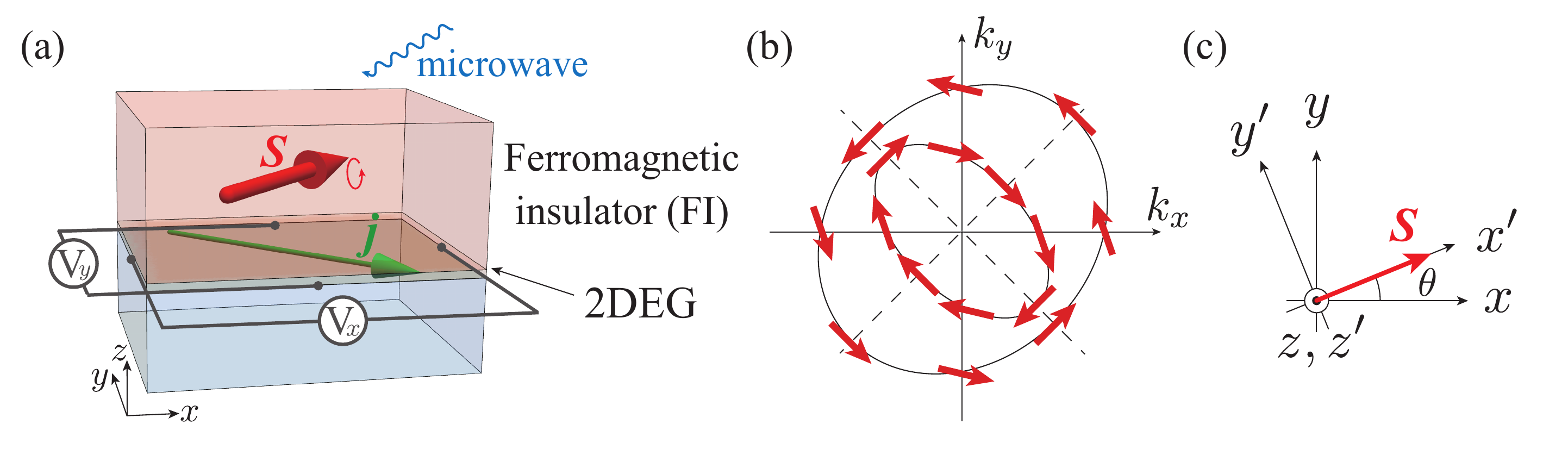}
\caption{(a) A junction system considered in our study. The red arrow, $\bm{S}$, indicates the spontaneous spin polarization of the ferromagnetic insulator (FI), which induces spin precession under microwave irradiation. The green arrow, $\bm{j}$, represents the current density generated by the inverse Rashba-Edelstein effect. (b) The schematic picture of the spin-split Fermi surface for $\alpha/\beta\sim 2$. Red arrows represent the spin polarizations of the energy eigenstates near the Fermi energy. (c) The coordinate transformation for the spin-wave approximation.}
\label{fig:setup}
\end{figure}

We consider a junction composed of 2DEG and FI as shown in Fig.~\ref{fig:setup}(a).
The Hamiltonian of 2DEG is written as $H_{\rm sys} = H_{\rm kin}+{\cal H}_{\rm imp}$, where $H_{\rm kin}$ and $H_{\rm imp}$ describe the kinetic energy and the impurity scattering, respectively.
The former is given as
\begin{align}
H_{\rm kin}&=\sum_{\bm{k},\sigma,\sigma'}
c^{\dagger}_{\bm{k}\sigma} (\hat{h}_{\bm{k}})_{\sigma\sigma'}
c_{\bm{k}\sigma'} ,\\
\hat{h}_{\bm{k}}&=\xi_{\bm k} \hat{I}+\alpha(k_y\hat{\sigma}_x-k_x\hat{\sigma}_y)+\beta(k_x\hat{\sigma}_x-k_y\hat{\sigma}_y).
\end{align}
Here, $c_{{\bm k}\sigma}$ is an annihilation operator of conduction electrons with a wavenumber ${\bm k}=(k_x,k_y)$ and a spin $\sigma$ ($=\uparrow,\downarrow$), $\xi_{\bm k} = \hbar^2{\bm k}^2/2 m^*-\mu$ is the kinetic energy measured from the chemical potential $\mu$, $m^*$ is the effective mass, $\hat{I}$ is a $2\times 2$ identity matrix, and $\hat{\bm \sigma} = (\hat{\sigma}_x,\hat{\sigma}_y,\hat{\sigma}_z)$ denotes the Pauli matrices.
The strengths of the Rashba and Dresselhaus spin-orbit interactions are denoted by $\alpha$ and $\beta$, respectively.
The $2\times 2$ matrix $\hat{h}_{\bm{k}}$ can be rewritten as $\hat{h}_{\bm k} =\xi_{\bm k} \hat{I} - {\bm h}_{\rm eff}({\bm k}) \cdot \hat{\bm \sigma}$, where ${\bm h}_{\rm eff}({\bm k})$ is
an effective Zeeman field acting on the conduction electrons, which is defined as
${\bm h}_{\rm eff}(\varphi) \simeq k_{\rm F}(-\alpha\sin\varphi -\beta\cos\varphi, \alpha\cos\varphi +\beta\sin\varphi)$ using the polar representation $\hat{\bm k}=(\cos \varphi,\sin \varphi)$.
The quantization axis of the electron spin depends on the electron wavenumber. As an illustrative example, we show the spin polarization near the Fermi surface for $\alpha/\beta = 2$ in Fig.~\ref{fig:setup}(b).
The Hamiltonian for impurity scattering due to
a short-range potential $v({\bm r}) = v_0 \delta({\bm r})$ is given as
\begin{align}
 H_{\rm imp} &= \frac{v_0}{{\cal A}}\sum_{{\bm k},{\bm q},\sigma} \rho_{\rm imp}({\bm q})c^\dagger_{{\bm k}+{\bm q}\sigma} c_{{\bm k}\sigma} , 
 \label{eq:Himp2d}
\end{align}
where ${\cal A}$ is an area of 2DEG,
$\rho_{\rm imp}({\bm q}) = \sum_i e^{-i{\bm q}\cdot {\bm R}_i}$ and ${\bm R}_i$ denote the position of the impurity.
By employing the Born approximation, the finite-temperature Green's function can be expressed as~\cite{Yama2021}
\begin{align}
& \hat{g}({\bm k},i\omega_m) = \frac{(i \hbar\omega_m  -\xi_{\bm k}+i \Gamma {\rm sgn}(\omega_m)/2)\hat{I} - {\bm h}_{\rm eff}\cdot \hat{\bm \sigma}}{\prod_{\nu=\pm} (i\hbar\omega_m-E_{\bm k}^\nu +i\Gamma {\rm sgn}(\omega_m)/2)},
\label{eq:gkwimp}
\end{align}
where $E_{\bm k}^{\pm} = \xi_{\bm k} \pm |{\bm h}_{\rm eff}(\varphi)|$ is the spin-dependent electron dispersion, $\Gamma = 2\pi n_i v_0^2 D_{\rm n}(\epsilon_{\rm F})$ is level broadening, and $n_i$ is the impurity concentration.

We introduce the proximity-induced exchange coupling between the FI and 2DEG explained in Sec.~\ref{sec:MicroscopicDescription}.
We employ the model for a clean interface (see Sec.~\ref{sec:MicroscopicDescription})\footnote{It can be shown that the spin injection rate does not depend on the in-plane orientation of the magnetic field for a dirty interface.}.
Furthermore, we assume that the spin in FI points in the $xy$ plane and define a new coordinate $x'y'$ to make the $x'$ direction align with the spin direction as shown in Fig.~\ref{fig:setup}(c).
Using the Holstein-Primakoff transformation explained in Sec.~\ref{sec:MicroscopicDescription}, the Hamiltonian for the proximity-induced exchange coupling is given as
\begin{align}
H_{\rm ex} &= \sum_{\bm k} \sqrt{2S_0}(J b_{\bm{k}} s^{x'-}_{\bm{k}}+ {\rm h.c.}). \label{eq:Hint2}
\end{align}
The spin ladder operator $s^{x'\pm}_{\bm{k}}=s_{\pm \bm k}^{y'}\pm i s_{\pm \bm k}^{z'}$ of 2DEG is defined in this new coordinate as
\begin{align}
\left(\begin{array}{c} s_{\bm q}^{x'} \\ s_{\bm q}^{y'}  \\ s_{\bm q}^{z'}\end{array} \right)
&= \left(\begin{array}{ccc}
\cos \theta & \sin \theta & 0 \\
-\sin \theta & \cos \theta & 0 \\
0 & 0 & 1 \end{array} \right) \left(\begin{array}{c} s_{\bm q}^{x} \\ s_{\bm q}^{y}  \\ s_{\bm q}^{z}\end{array} \right), 
\end{align}
where the spin operators, $s_{\bm q}^{a}$ ($a=x,y,z$), are defined in Eq.~(\ref{sqdef}).
In the following calculation, we consider second-order perturbation with respect to $H_{\rm ex}$, assuming that the spin-splitting energy is much smaller than the Fermi energy and much larger than the impurity scattering rate and temperature.

\subsection{Enhancement of Gilbert damping}

\begin{figure}[tb]
\centering
\includegraphics[width=150mm]{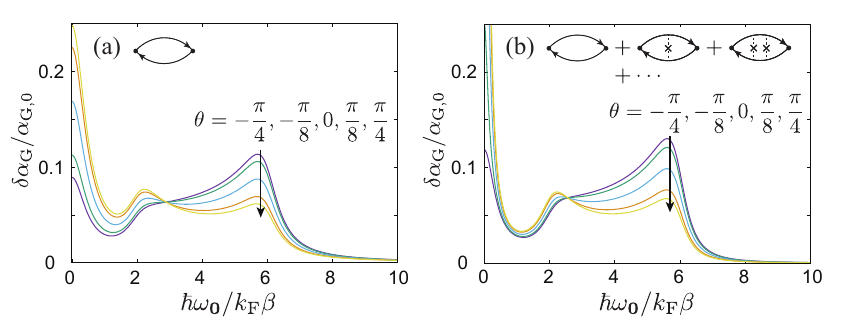}
\caption{Modulation of the Gilbert damping (a) without vertex corrections and (b) with vertex corrections is plotted for $\alpha/\beta = 2$ as a function of the FMR frequency $\omega_{\bm 0}$. 
The inset in each graph indicates the corresponding Feynman diagram for the spin susceptiblity, where the cross indicates the impurity scattering.}
\label{fig:BetheSalpeter}
\end{figure}

As described in Sec.~\ref{sec:GilbertDamping}, the enhancement of the Gilbert damping, $\delta \alpha_{\mathrm{G}}$, is obtained by the calculation of the spin susceptibility in 2DEG.
In the simplest approximation, the spin susceptibility is expressed by the Feynman diagram shown in the inset of Fig.~\ref{fig:BetheSalpeter}(a), where the solid line denotes the electron propagator in the Born approximation given in Eq.~(\ref{eq:gkwimp}).
Then, the spin susceptibility is written as
\begin{align}
\chi({\bm 0},i\omega_n) 
&= \frac{k_{\rm B}T}{4{\cal A}}
\sum_{{\bm k},i\omega_m} {\rm Tr} \Biggl[ \hat{\sigma}^{x'-}
\hat{g}({\bm k},i\omega_m)  \hat{\sigma}^{x'+}
\hat{g}({\bm k},i\omega_m+i\omega_n) \Biggr]\label{eq:self_Sigma}.
\end{align}
where $\hat{\sigma}^{x'\pm} = \hat{\sigma}^{y'}\pm i \hat{\sigma}^{z'} = -\sin \theta \, \hat{\sigma}_x + \cos \theta \, \hat{\sigma}_y \pm i \hat{\sigma}_z$ is a $2\times 2$ matrix.
By a standard procedure of the Green's function method~\cite{Bruus2004}, the retarded spin susceptibility is obtained by $\chi^R({\bm 0},\omega) = \chi({\bm 0},i\omega\rightarrow \omega + i\delta)$ and the enhancement of the Gilbert damping can be expressed as \cite{Yama2021}
\begin{align}
\delta\alpha_{\rm G} & = \delta \alpha_{{\rm G},1} + \delta \alpha_{{\rm G},2}+ \delta \alpha_{{\rm G},3} \\
\delta \alpha_{{\rm G},1} &= \alpha_{{\rm G},0} \int_0^{2\pi} \! \frac{d\varphi}{2\pi} \, 
\frac{\Gamma 
k_{\rm F}\beta/\pi}{\hbar^2 \omega_{\bm 0}^2 + \Gamma^2} \frac{1-(\hat{\bm h}_{\rm eff}\cdot \hat{\bm m})^2}{2}, 
\label{eq:result1} \\
\delta \alpha_{{\rm G},2} &= \alpha_{{\rm G},0} \int_0^{2\pi} \! \frac{d\varphi}{2\pi} \, \frac{\Gamma 
k_{\rm F}\beta/\pi}{(\hbar \omega_{\bm 0}-2h_{\rm eff}(\varphi))^2 + \Gamma^2}\frac{(1+\hat{\bm h}_{\rm eff}\cdot \hat{\bm m})^2}{4}, 
\label{eq:result2} \\
\delta \alpha_{{\rm G},3} &= \alpha_{{\rm G},0} \int_0^{2\pi} \! \frac{d\varphi}{2\pi} \, \frac{\Gamma 
k_{\rm F}\beta/\pi}{(\hbar \omega_{\bm 0}+2h_{\rm eff}(\varphi))^2 + \Gamma^2} \frac{(1-\hat{\bm h}_{\rm eff}\cdot \hat{\bm m})^2}{4},
\label{eq:result3}
\end{align}
where $\alpha_{{\rm G},0} = 2 \pi S_{0}|J|^2 {\cal A} D(\epsilon_{\rm F})/k_{\rm F}\beta$ 
is a dimensionless parameter, $\hat{\bm m}=(\cos \theta,\sin \theta)$ denotes the direction of the spin in the FI, and $h_{\rm eff}(\varphi)=|{\bm h}_{\rm eff}(\varphi)|$.
The first contribution $\delta \alpha_{{\rm G},1}$ comes from elastic spin flipping of conduction electrons caused by the transverse component of the effective magnetic field ${\bm h}_{\rm eff}$ via the interfacial exchange coupling.
In fact, $\delta \alpha_{{\rm G},1}$ has a maximum when ${\bm h}_{\rm eff}$ is perpendicular to $\hat{\bm m}$.
The second and third contributions, $\delta \alpha_{{\rm G},2}$ and $\delta \alpha_{{\rm G},3}$, originate from the magnon absorption and emission, respectively.
Fig.~\ref{fig:BetheSalpeter}(a) illustrates the enhancement of the Gilbert damping as a function of the resonant frequency $\omega_{\bm 0}$ for $\Gamma/k_{\rm F}\beta = 0.5$ and $\alpha/\beta = 2$.
The Gilbert damping clearly depends on the spin orientation, $\theta$ and has a peak structure at $\omega_{\bm 0} = 0$ and a broad structure in the range of $2k_{\rm F}\beta \le \hbar \omega_{\bm 0} \le 6 k_{\rm F}\beta$.
The former peak corresponds to elastic spin flipping described by $\delta \alpha_{{\rm G},1}$, while the latter structure is produced by $\delta \alpha_{{\rm G},2}$ through the variation in the spin splitting energy $2h_{\rm eff}(\varphi)$.
We note that the third contribution, $\delta \alpha_{{\rm G},3}$, is not important in usual FMR experiments.

For a more accurate calculation, we need to consider the vertex corrections so that the Ward-Takahashi relation holds~\cite{Yama2023}.
For the self-energy in the Born approximation, the spin susceptibility with the vertex corrections is expressed by the ladder diagram~\cite{Fulde1968,Bruus2004,Akkermans2007} as shown in the inset of Fig.~\ref{fig:BetheSalpeter}(b). 
The main graph of Fig.~\ref{fig:BetheSalpeter}(b) shows the enhancement of the Gilbert damping calculated by considering the vertex corrections for the same parameters as Fig.~\ref{fig:BetheSalpeter}(a).
While the qualitative features are unchanged, the peak at $\omega_{\bm 0} = 0$, which originates from elastic spin flipping, is remarkably enhanced when the vertex corrections are considered.
The change after taking into account the vertex corrections becomes more significant for $\alpha/\beta \simeq 1$~\cite{Yama2023}.

\subsection{Inverse Rashba-Edelstein effect (IREE)}
\label{sec:IREE}

In 2DEG, the non-equilibrium spin accumulation is converted into a charge current via the inverse Rashba-Edelstein effect (IREE)~\cite{Ganichev2002, Sanchez2013, Shen2014a, Suzuki2023, Yama2023b, Hosokawa2024}.
Therefore, spin pumping by microwave irradiation can generate a voltage (a charge current) in 2DEG through the IREE (see Fig.~\ref{fig:setup}(a)).
In recent years, experimental studies on the IREE using spin pumping have become prevalent across various systems, such as Ag/Bi~\cite{Sanchez2013, Nomura2015, Sangiao2015, Zhang2015, Matsushima2017}, topological insulators~\cite{Shiomi2014, Sanchez2016, Wang2016, Song2016, Mendes2017, Sun2019, Singh2020, Dey2021, He2021, Zhang2016}, STO~\cite{Lesne2016, Song2017, Vaz2019, Noel2020, Ohya2020, Bruneel2020, To2021, Trier2022}, and semiconductors~\cite{Chen2016, Oyarzun2016}. 
A few theoretical studies related to the IREE driven by spin pumping have been conducted for two-dimensional electron systems~\cite{Tolle2017,Dey2018}.
However, these theoretical studies are based on a static interaction across the junctions and neglect dynamic processes such as magnon absorption and emission.

The theoretical framework for taking into account dynamic processes has been proposed in Ref.~\cite{Yama2023b}.
In this study, 2DEG with both the Rashba and Dresselhaus spin-orbit interactions was considered in the regime of $k_{\rm F}\alpha,k_{\rm F}\beta \gg \Gamma$.
The distribution function of conduction electrons in 2DEG was expressed for a uniform steady state as $f({\bm k},\gamma)$, where $\gamma$ is an index of the spin-polarized bands~\cite{Suzuki2023}.
Then, the Boltzmann equation is described as
\begin{align}
0 = \frac{\partial f}{\partial t}\biggl{|}_{\rm pump}+\frac{\partial f}{\partial t}\biggl{|}_{\rm imp} , \label{eq:0ecoll}
\end{align}
where $\partial f/\partial t|_{\rm pump}$ is a collision term due to spin injection from the FI into the 2DEG through the interface, and $\partial f/\partial t|_{\rm imp}$ is a collision term due to impurity scattering.
Explicit expressions of the collision terms are given in \ref{app:CollisionTerm}.
For the linear response to external driving, the distribution function can be approximated as~\cite{Ziman1960, Lundstrom2000}
\begin{align}
f(\bm{k},\gamma) \simeq f_0(E^{\gamma}_{\bm{k}})-\frac{\partial f_0(E^{\gamma}_{\bm{k}})}{\partial E^{\gamma}_{\bm{k}}} \delta\mu(\varphi,\gamma).
\label{eq:fexpan}
\end{align}
where $E^{\gamma}_{\bm{k}}$ is an energy dispersion of electrons in 2DEG, $f_0(\epsilon) =(\exp[ (\epsilon-\mu)/k_{\rm B}T]+1)^{-1}$ is the Fermi distribution function, and $\delta \mu(\varphi,\gamma)$ is a chemical potential shift in the direction of $\varphi$.
Combining these expressions of the collision terms with the Boltzmann equation, we can calculate the nonequilibrium distribution function $f({\bm k},\gamma)$ (for a detailed calculation, see Ref.~\citeonline{Yama2023b}).
Note that our formulation does not need the concept of spin current.

\begin{figure}[tb]
\centering
\includegraphics[width=150mm]{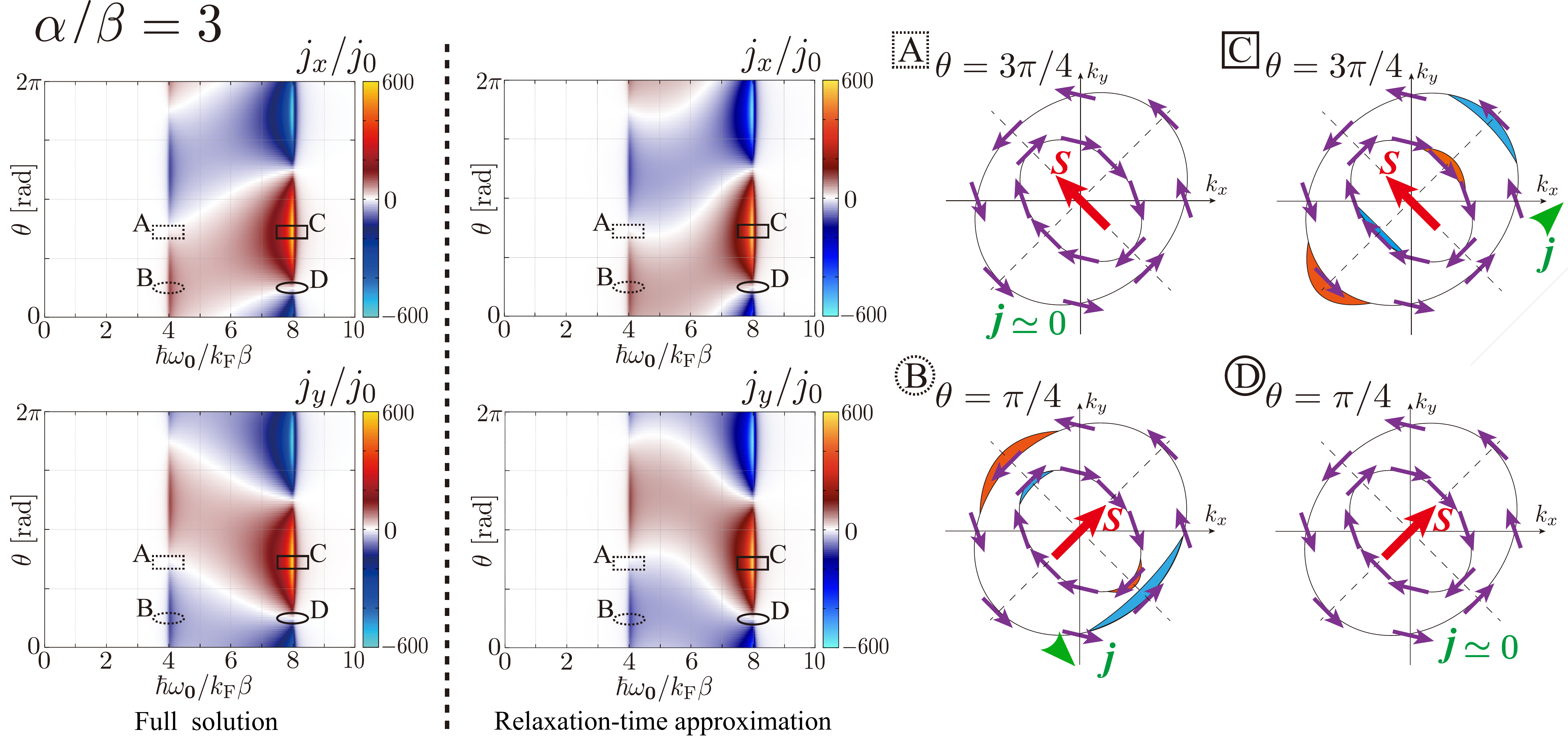}
\caption{Left contour graphs: The charge currents in the $x$ and $y$ directions as a function of the FMR frequency and the orientation of the spin in FI. Middle contour graphs: The charge currents obtained by the relaxation-time approximation. Right four  pictures: Schematic of the Fermi surface modulation and the direction of the induced current at four points, A, B, C, and D, in the contour graphs.}
\label{fig:2DEGIREE}
\end{figure}

The two contour plots on the left of Fig.~\ref{fig:2DEGIREE} show the current density as functions of the FMR frequency and the azimuth angle of the spin in FI for $\alpha/\beta=3$.
Due to the distribution of $h_{\rm eff}(\varphi)$, the current densities exhibit large values across a wide range of $4 k_{\rm F}\beta \lesssim \hbar \omega_{\bm 0} \lesssim 8 k_{\rm F}\beta$. 
Near the lower boundary at $\hbar \omega_{\bm 0}/k_{\rm F}\beta = 4$, the current amplitude vanishes at $\theta=3\pi/4$ (point A in the contour plot) and takes a maximum at $\theta=\pi/4$ (point B), where the current flows along the $(1,-1)$ direction.
On the other hand, near the higher boundary at $\hbar \omega_{\bm 0}/k_{\rm F}\beta = 8$, the current amplitude takes a maximum at $\theta=3\pi/4$ (point C), where the current flows along the $(1,1)$ direction, and vanishes at $\theta=\pi/4$ (point D).
The Fermi surface modulations at the corresponding points are schematically shown in the right four pictures of Fig.~\ref{fig:2DEGIREE}.
It is remarkable that the full solution of the Boltzmann equation is needed for an accurate description of the IREE.
If we employ the relaxation-time approximation~\cite{Silsbee2004,Yama2024a}, the results are changed significantly as shown in the middle contour plots of Fig.~\ref{fig:2DEGIREE}.

\subsection{Related Phenomena}

Spin-charge conversion by IREE can also be driven by a thermal gradient in a junction composed of 2DEG and FI. In fact, the current generation by thermally induced magnon spin current has been observed in the 2DEG at the EuO--KTaO${}_3$ heterostructure~\cite{Zhang19}.
The interplay of the Rashba-Edelstein effect (REE) and IREE also produces a characteristic magnetoresistance termed the Rashba-Edelstein magnetoresistance~\cite{Nakayama2016,Nakayama2017,Nakayama2018,Thompson2020}. 
These phenomena can be discussed by the Boltzmann equation with collision terms in a similar way as Sec.~\ref{sec:IREE} \cite{Hosokawa2024,Yama2025}.

\section{Spin pumping into TMDC}
\label{sec_SP_TMDC}

\begin{figure}
\begin{center}
\includegraphics[width=1.0\hsize]{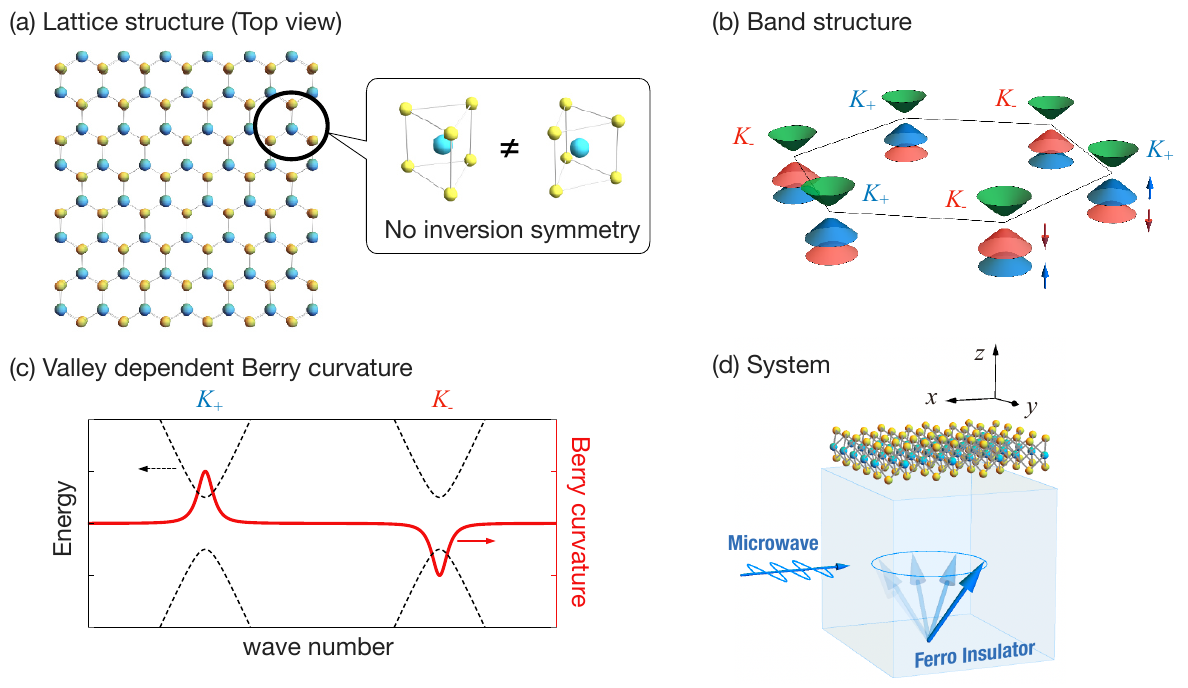}
\end{center}
\caption{Distinctive properties of monolayer TMDCs.
(a) When viewed from above, the lattice structure forms a honeycomb lattice. The unit cell lacks a inversion center, breaking spatial inversion symmetry.
(b) Schematic picture of the band structure of monolayer TMDCs. A Zeeman-type spin splitting occurs due to the strong SOC of the transition metals. Because of the mirror symmetry of the crystal structure, the $z$-component of the spin becomes a conserved quantity up to first order of wave number in the model Hamiltonian.
(c) Schematic picture of valley-dependent Berry curvature. The Berry curvature exhibits peaks in both $K_+$ and $K_-$ valleys, with their signs being opposite.
(d) A schematic illustration of the system where the FI is attached to the TMDC. Under microwave irradiation, spin pumping occurs at the interface between the FI and the TMDC.}
\label{fig_TMDC_properties}
\end{figure}

Monolayer TMDCs have a unique band structure that has attracted considerable attention. When viewed from the top, the lattice forms a honeycomb structure, with each unit cell consisting of one transition metal atom and two chalcogen atoms, as shown in Fig.\ \ref{fig_TMDC_properties}\ (a). The unit cell lacks an inversion center, resulting in broken spatial inversion symmetry. The band extrema are located at the $K_+$ and $K_-$ points, the corners of the Brillouin zone \cite{Mak2010,Splendiani2010}, as shown in Fig.\ \ref{fig_TMDC_properties}\ (b). Due to the broken spatial inversion symmetry and the strong SOC of the transition metal atoms, the band structure exhibits spin–valley coupling (SVC) \cite{Xiao2012}. Furthermore, the broken inversion symmetry gives rise to a finite valley-dependent Berry curvature \cite{xiaoValleyContrastingPhysicsGraphene2007}, as shown in Fig.\ \ref{fig_TMDC_properties}\ (c).

The recent accelerated advancements in the fabrication techniques of TMDC devices have significantly expanded our understanding of valley physics, including valley-dependent circular dichroism \cite{caoValleyselectiveCircularDichroism2012,makControlValleyPolarization2012,zengValleyPolarizationMoS2012,sallenRobustOpticalEmission2012,wuElectricalTuningValley2013,zhaoEnhancedValleySplitting2017,zhongVanWaalsEngineering2017,seylerValleyManipulationOptically2018,nordenGiantValleySplitting2019}, the valley Hall effect \cite{makValleyHallEffect2014,leeElectricalControlValley2016a,ubrigMicroscopicOriginValley2017,wuIntrinsicValleyHall2019,hungDirectObservationValleycoupled2019a}, and valley-dependent spin injection via spin-polarized charge injection \cite{yeElectricalGenerationControl2016}.
All of these experiments have harnessed charge excitations induced by electric fields and optical irradiation.
In contrast, SVC offers a potential pathway to access the valley degrees of freedom through spin excitation.

In this section, we theoretically investigate the spin pumping into a monolayer TMDC \cite{Ominato2020b}.
Figure \ref{fig_TMDC_properties}\ (d) shows a schematic of the system, where an FI is attached to a monolayer TMDC. 
Microwave irradiation is then applied to the system, triggering FMR.
Through spin-transfer processes mediated by proximity exchange coupling at the interface, the FMR excites electron spins in the monolayer TMDC.
Our findings reveal that the interplay between SVC and proximity exchange coupling leads to valley-selective spin excitation.

\subsection{Model}

We consider an FI/TMDC hybrid system, where the FI weakly perturbs the band structure of the TMDC, and the energy bands of the FI are absent in the energy range considered here.
The Hamiltonian $\Hsys = \sum_{\vk,\eta} \varepsilon_{\vk\eta} c^\dagger_{\vk\eta} c_{\vk\eta}$ describes the electronic states of the monolayer TMDC,
where $c^\dagger_{\vk\eta}$ ($c_{\vk\eta}$) is the electron creation (annihilation) operator with eigenenergy $\varepsilon_{\vk\eta}$ and quantum number $\eta = (n, \tau, s)$, where $n = \pm$, $\tau = \pm$, and $s = \pm$ denote the band, valley, and spin indices, respectively.
The eigenenergies and eigenstates are obtained by diagonalizing the effective Hamiltonian around the $K_+$ and $K_-$ points \cite{Xiao2012}
\begin{align}
    \hat{h}_{\vk}=
        &\hbar v
        \left(
            \tau k_x\hat{\sigma}^x+
            k_y\hat{\sigma}^y
        \right)
        +\frac{\Delta}{2}\hat{\sigma}^z
        -\tau s\lambda\frac{\hat{\sigma}^z-1}{2},
    \label{eq_effective}
\end{align}
where $v$ is the velocity, $\Delta$ is the energy gap, $\lambda$ is the spin splitting at the valence-band top caused by SOC, and $\hat{\bm{\sigma}}$ denotes the Pauli matrices acting on the orbital degrees of freedom.
These parameters are fitted from first-principles calculations
\cite{zhuGiantSpinorbitinducedSpin2011,cheiwchanchamnangijQuasiparticleBandStructure2012,kormanyosMonolayerMoSTrigonal2013,liuThreebandTightbindingModel2013,kangBandOffsetsHeterostructures2013a,kormanyosTheoryTwodimensionalTransition2015,echeverrySplittingBrightDark2016}.
As mentioned in Sec.~\ref{sec:MicroscopicDescription}, the proximity exchange coupling consists of two terms, $\Hex=\HZ+\HT$.
$\HZ$ modulates the spin splitting, and $\HT$ describes spin transfer at the interface.
In general, $\HT$ includes both intravalley and intervalley spin-transfer processes.
Assuming that the characteristic length scale of the interface roughness exceeds the lattice constant of the TMDC, the intervalley process can be neglected.
Consequently, each valley can be treated independently.

\subsection{Interface spin current}

We obtain the following analytical expression for the spin current
\begin{align}
    \la I_{\mathrm{s}}\ra = I_{\mathrm{s}}^{K_+} + I_{\mathrm{s}}^{K_-},
\end{align}
where we introduce the valley-resolved spin current
\begin{align}
    I_{\mathrm{s}}^{K_\tau}
    =
    \frac{|J|^2\mathcal{A}(S_0\gamma_{\mathrm{g}}h_{\mathrm{ac}})^2}{(\omega-\omega_{\bm{0}})^2+\alpha_{\mathrm{G}}^2\omega_{\bm{0}}^2}
    \sum_{\vq}\mathrm{Im}\,\chi^R_{\tau}(\vq,\omega),
    \label{eq_spin_current}
\end{align}
with the local spin susceptibility for each valley.
The imaginary part of the local spin susceptibility is given by
\begin{align}
    \sum_{\vq}\mathrm{Im}\,\chi^R_{\tau}(\vq,\omega)
        = -2\pi\hbar\omega
            \int d\varepsilon
            \left(
                -\frac{\partial f(\varepsilon)}{\partial \varepsilon}
            \right)
            D_{\tau,+}(\varepsilon) D_{\tau,-}(\varepsilon)
            \left(
                1 + \frac{Z_{\tau,+} Z_{\tau,-}}{|\varepsilon - E_{\tau,+}|\, |\varepsilon - E_{\tau,-}|}
            \right),
    \label{eq_susceptibility}
\end{align}
where $f(\varepsilon) = 1/\left(e^{(\varepsilon - \mu)/\kbt} + 1\right)$ is the Fermi distribution function with chemical potential $\mu$ and temperature $T$, and
$D_{\tau,s}(\varepsilon)$ is the density of states per unit area
\begin{align}
    D_{\tau,s}(\varepsilon) =
        \frac{1}{2\pi(\hbar v)^2}
        |\varepsilon - E_{\tau,s}|
        \, \theta(|\varepsilon - E_{\tau,s}| - Z_{\tau,s}),
\end{align}
with $E_{\tau,s} = s(\tau\lambda/2 - JS_0)$ and $Z_{\tau,s} = \Delta/2 - \tau s \lambda/2$.
At zero temperature, the spin current is finite when the product of the up- and down-spin density of states in each valley is finite at the Fermi level, as shown in Eqs.~(\ref{eq_spin_current}) and (\ref{eq_susceptibility}).

\subsection{Numerical results}

\begin{figure*}
\begin{center}
\includegraphics[width=0.9\hsize]{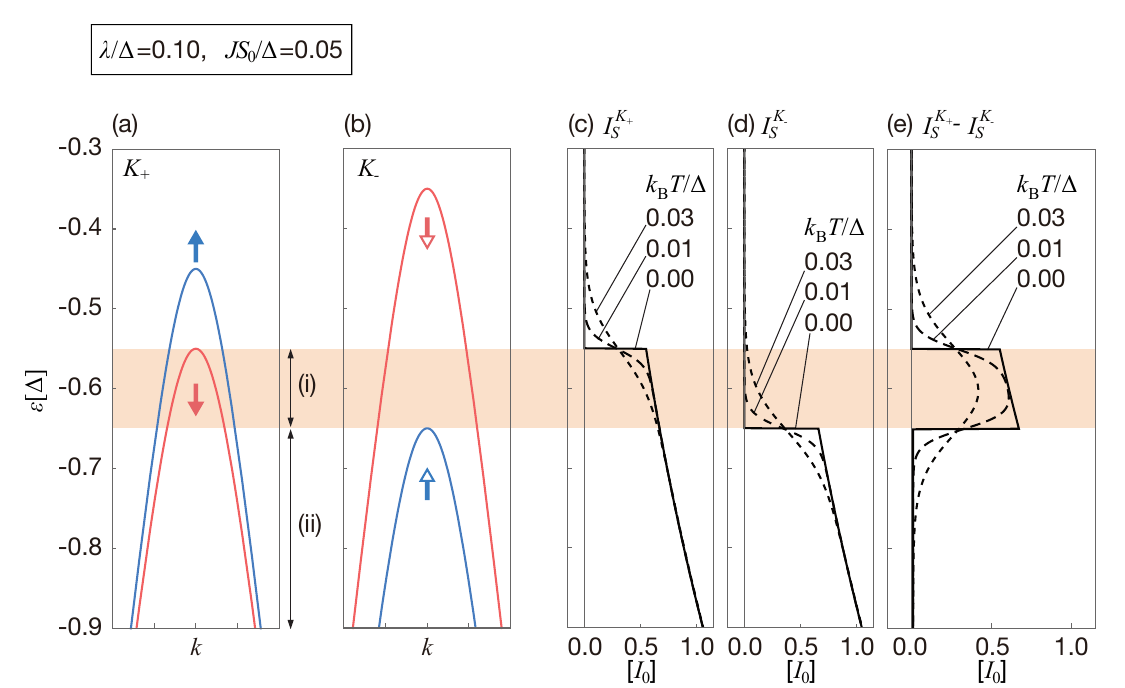}
\end{center}
\caption{
The panels (a) and (b) show the valence bands in the $K_+$ and $K_-$ valleys, respectively. The panels (c) and (d) show the spin current at several temperatures in the $K_+$ and $K_-$ valleys, respectively. The panel (e) shows the valley-polarized spin current at several temperatures. The units of the spin current are given by $I_0=\frac{|J|^2\mathcal{A}(S_0\gamma_{\mathrm{g}}h_{\mathrm{ac}})^2\hbar\omega}{(\omega-\omega_{\bm{0}})^2+\alpha_{\mathrm{G}}^2\omega_{\bm{0}}^2}\frac{\Delta^2}{(\hbar v)^4}$.}
\label{fig_spin_current}
\end{figure*}

A fundamental result derived from the above expressions is the feasibility of generating a valley-polarized spin current (VPSC).
This is because the valley degeneracy is lifted by the proximity exchange coupling, allowing the spin current in each valley to differ.
To characterize the valley polarization of the spin current, we define the VPSC as $I_{\mathrm{s}}^{K_+} - I_{\mathrm{s}}^{K_-}$.
We show that, with appropriate carrier doping, spins can be valley-selectively excited, resulting in a completely VPSC.
Panels (a) and (b) of Fig.\ \ref{fig_spin_current} show the valence bands in the $K_+$ and $K_-$ valleys, respectively.
We set the parameters as $\lambda/\Delta = 0.10$ and $JS_0/\Delta = 0.05$, consistent with values from first-principles calculations
\cite{qiGiantTunableValley2015a,zhangLargeSpinValleyPolarization2016,liangMagneticProximityEffect2017,xuLargeValleySplitting2018}.
Panels (c), (d), and (e) of Fig.\ \ref{fig_spin_current} show the spin current in the $K_+$ valley, the spin current in the $K_-$ valley, and the VPSC, respectively.
In energy region (i), the spin current at zero temperature is finite only in the $K_+$ valley, indicating valley-selective spin excitation and the generation of a completely VPSC.
This condition remains valid even at finite temperatures provided that the spin splitting due to proximity exchange coupling is much greater than thermal broadening, i.e., $JS_0 \gg \kbt$.
In energy region (ii), however, the spin current is finite in both valleys, leading to an almost zero VPSC and suppressed valley selectivity.
We also note that a small spin splitting exists in the conduction band \cite{zhuGiantSpinorbitinducedSpin2011,cheiwchanchamnangijQuasiparticleBandStructure2012,kormanyosMonolayerMoSTrigonal2013,liuThreebandTightbindingModel2013}, which is omitted in our model Hamiltonian.
Thus, valley-selective spin excitation is also feasible in the conduction band.

The present review concentrates on monolayer TMDCs, while extending the theoretical framework to multilayers remains an open and important problem. Spin pumping into multilayer TMDCs has already been demonstrated experimentally \cite{husain2018,Husain2022}, highlighting the significance of this research avenue. As the layer number increases, the band gap evolves from direct to indirect \cite{Mak2010,Splendiani2010}. Symmetry analysis indicates that inversion symmetry is absent in odd layers, which lifts spin degeneracy and enables spin–valley coupling, whereas even layers preserve inversion symmetry and remain spin-degenerate \cite{Xiao2012}. Odd-layer systems are therefore expected to exhibit physics analogous to that discussed in this review, and a detailed theoretical investigation will be pursued in future work.

\section{Other Applications}
\label{sec:perspectives}

In this review, we have introduced spin pumping and related topics, with a particular focus on superconductors, 2DEG, and TMDCs. Spin pumping can also be used as a probe of the spin-split band structure in various materials. Theoretical studies on spin pumping into Landau levels of graphene under a magnetic field~\cite{Ominato2020a}, twisted bilayer graphene~\cite{Haddad2023}, anisotropic Dirac electrons~\cite{Funato2022}, Andreev bound states in unconventional superconductors~\cite{Silaev2020b,Sun2023b}, and altermagnets~\cite{Sun2023} have been conducted so far.
Spin pumping has been studied theoretically as a probe of spin excitations in interacting electron systems such as Kondo impurities on a surface~\cite{yamamoto2021}, non-Fermi liquid states~\cite{Zhang2025}, 
carbon nanotubes~\cite{Fukuzawa2023}, one-dimensional spin chains~\cite{Furuya2024}, and spin nematic states~\cite{Ishikawa2024}.
It is anticipated that these theoretical predictions will be confirmed in future experiments, which may provide deeper insight into the spin pumping technique.

Spin pumping has also been extended to organic semiconductors, attractive materials due to their potential for long spin lifetimes and chemical tunability~\cite{Gu2024}. A pioneering demonstration of spin-charge conversion via the ISHE was achieved in a solution-processed conducting polymer~\cite{Ando2013}. A key advantage of these molecular systems is the ability to engineer their properties. Indeed, it was subsequently shown that the ISHE can be controlled by chemically tuning the spin-orbit coupling of the molecules~\cite{Sun2016}. The potential of this approach has been further underscored by reports of giant spin pumping at an optimized ferromagnet-organic interface, with efficiencies comparable to those of heavy metals~\cite{Manoj2021}. However, reliably quantifying the ISHE remains a central challenge, as the genuine signal is often obscured by parasitic thermoelectric and spin rectification effects, necessitating careful experimental methodologies to ensure accurate interpretation~\cite{Qaid2020, Skalski2022}.

Recently, orbital current, i.e., the flow of orbital angular momentum, has attracted considerable attention in the field of spintronics and has been extensively studied~\cite{Bernevig2005,Go2021,Johansson2024}. Among the various phenomena associated with orbital current, orbital pumping, which is an extension of spin pumping to orbital angular momentum, has been investigated both experimentally and theoretically~\cite{Hayashi2024,Go2025}. A detailed investigation of orbital pumping based on microscopic models remains an important direction for future research.

\section{Summary}
\label{sec_summary}

In this review, we have provided an overview of recent theoretical advances in the study of spin transport phenomena, focusing on systems in which ferromagnetic resonance (FMR) modulation serves as a probe of spin-dependent processes at interfaces between two-dimensional materials and magnetic layers.  
We have shown that changes in the FMR linewidth, arising from spin pumping across the interface, can be used to investigate spin transport properties in low-dimensional systems.

We first considered spin transport into two-dimensional superconductors, showing that FMR measurements offer a means to detect superconducting pairing symmetries by analyzing the enhancement of Gilbert damping.  
In particular, we emphasized that FMR modulation provides a sensitive probe of spin excitation processes in superconducting states, offering a complementary approach to conventional spin probes and enabling more comprehensive access to spin-related information.  
This approach opens up opportunities to explore unconventional superconductivity using FMR measurements.

We then discussed spin transport into two-dimensional electron gases formed at semiconductor heterostructures, where Rashba and Dresselhaus spin–orbit interactions play key roles.  
In such systems, the ability to tune the spin texture via external electric fields enables control over spin dynamics and transport.  
We have shown that spin pumping into these systems leads to distinct modifications of the FMR signal, providing access to information on the spin–orbit-coupled electronic structure and mechanisms of spin current generation.

Finally, we examined spin transport phenomena in monolayer transition metal dichalcogenides (TMDCs).
These materials, characterized by strong spin–orbit coupling and broken inversion symmetry, exhibit spin–valley coupling and finite Berry curvature effects.
We have shown that spin pumping into these TMDCs can induce valley-selective spin excitations, offering new routes for controlling valley degrees of freedom via spin excitations.

In summary, the studies reviewed here highlight the versatility of FMR modulation as a probe of spin transport phenomena in two-dimensional systems.  
Our findings underscore the potential of two-dimensional superconductors, semiconductor heterostructures, and atomically thin materials as platforms for future spintronic devices, and point to promising directions for the further development of spin-based technologies in low-dimensional systems.

\section*{Acknowledgments}

The authors are grateful to J. Fujimoto, T. Funato, H. Funaki, K. Hosokawa, M. Kohda, Y. Suzuki, and Y. Kato for valuable discussions.
This work was supported by the National Natural Science Foundation of China (NSFC) under Grant No. 12374126, 
by the Priority Program of Chinese Academy of Sciences under Grant No. XDB28000000, Waseda University Grant for Special Research Projects (Grants No. 2025C-651 and No. 2025R-061), and by JSPS KAKENHI for Grants (Nos. JP21H01800, JP21H04565, JP23H01839, JP24K06951, JP24H00322, JP24H00853, JP24KJ0624, and JP25K07224) from MEXT, Japan.

\appendix

\section{Holstein-Primakoff transformation}
\label{appA}

Here, we introduce the Holstein-Primakoff transformation \cite{holstein1940field}. The localized spin $\vS_n$ is described as shown using the boson creation (annihilation) operator $b_n^\dagger$ ($b_n$)
\begin{align}
    &S^+_n=S^x_n+iS^y_n=\qty(2S_0-b_n^\dagger b_n)^{1/2}b_n, \\
    &S^-_n=S^x_n-iS^y_n=b_n^\dagger\qty(2S_0-b_n^\dagger b_n)^{1/2}, \\
    &S^z_n=S_0-b_n^\dagger b_n,
\end{align}
where we require $[b_n,b_m^\dagger]=\delta_{n,m}$ to ensure that $S_n^+$, $S_n^-$, and $S_n^z$ satisfy the commutation relation of angular momentum. The deviation of $\la S_n^z\ra$ from its maximum value $S_0$ is quantified using the boson particle number $\la b_n^\dagger b_n\ra$.
Assuming a small deviation such that $\la b_n^\dagger b_n\ra/S_0\ll1$, the ladder operators $S_n^\pm$ can be approximated by $S_n^+\simeq(2S_0)^{1/2}b_n$ and $S_n^-\simeq(2S_0)^{1/2}b_n^\dagger$. This is known as the spin-wave approximation.
Performing Fourier expansion, the boson operators are given by
\begin{align}
    &b_n^\dagger=\frac{1}{\sqrt{N_{\rm FI}}}\sum_{\vk}e^{-i\vk\cdot\vr_n}b_{\vk}^\dagger, \\
    &b_n=\frac{1}{\sqrt{N_{\rm FI}}}\sum_{\vk}e^{i\vk\cdot\vr_n}b_{\vk},
\end{align}
where $N_{\rm FI}$ is the number of sites.
The boson operators $b_{\vk}^\dagger$ and $b_{\vk}$ with wave vector $\vk$ satisfy $[b_{\vk},b_{\vk^\prime}^\dagger]=\delta_{\vk,\vk^\prime}$ and describe the collective excitation called magnon.
Substituting these approximate relations, the Hamiltonian of the FI, Eq.~(\ref{Heisenberg}), is modified as Eq.~\eqref{HFImagnon} in the leading order of $1/S_0$.

\section{Derivation of the spin current formula}
\label{appB}

\begin{figure}
\begin{center}
\includegraphics[width=0.5\hsize]{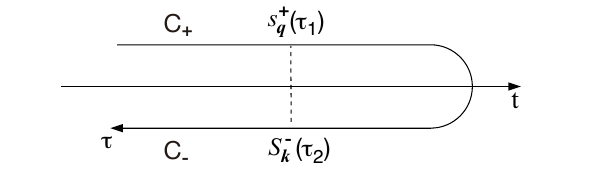}
\end{center}
\caption{
The Keldysh contour $C$ is componsed of the forward path $C_+$ running from $-\infty$ to $\infty$ and the backward path $C_-$ running from $\infty$ to $-\infty$. The time variables $\tau_1$ and $\tau_2$ have been put on $C_+$ and $C_-$, respectively.}
\label{fig_contour}
\end{figure}

To perform perturbation calculations, we implement the interaction picture and examine the time evolution along the Keldysh contour as shown in Fig.\ \ref{fig_contour}. Subsequently, the spin current is given by
\begin{align}
    \la \IS(\tau_1,\tau_2)\ra 
    =\mathrm{Re}\qty[\frac{i}{2}\sum_{\vk,\vq}J_{\vk,\vq}\la T_C\mathcal{S}_Cs^+_{\vq}(\tau_1)S_{\vk}^-(\tau_2)\ra],
\end{align}
where $T_C$ is the time ordering operator, $\mathcal{S}_C$ is given by
\begin{align}
    \mathcal{S}_C=T_C\exp\qty(-\frac{i}{\hbar}\int_Cd\tau\Hex(\tau)),
\end{align}
and we have put the time variable $\tau_1$ and $\tau_2$ on the forward path $C_+$ and the backward path $C_-$, respectively.
Expanding $\mathcal{S}_C$ and keeping the lowest-order term with respect to $\HT$, we obtain
\begin{align}
    \la \IS(\tau_1,\tau_2)\ra
    =\frac{1}{2\hbar}\int_C d\tau
    \sum_{\vk,\vq}|J_{\vk,\vq}|^2
    \mathrm{Re}
    \qty[
        \la T_Cs^+_{\vq}(\tau_1)s^-_{\vq}(\tau)\ra
        \la T_CS^+_{\vk}(\tau)S^-_{\vk}(\tau_2)\ra
    ].
\end{align}
Assuming a steady state and using real time representation, the spin current is given by
\begin{align}
    \la \IS\ra &
    =\frac{\hbar}{2}\int^{\infty}_{-\infty}dt\sum_{n,n^\prime}|J_{\vk,\vq}|^2
    \mathrm{Re}
    \big[
        \chi^{R}(\vq,t)G^<(\vk,-t) +
        \chi^{<}(\vq,t)G^A(\vk,-t)
        \big].
\end{align}
Performing Fourier transformation of the spin susceptibility, the spin current is given by
\begin{align}
    \la \IS\ra & =\frac{\hbar}{2}
    \int^{\infty}_{-\infty} \frac{d\omega}{2\pi}
    \sum_{\vk,\vq}|J_{\vk,\vq}|^2
    \mathrm{Re}
    \qty[
        \chi^{R}(\vq,\omega)G^<(\vk,\omega)+
        \chi^{<}(\vq,\omega)G^A(\vk,\omega)
        ].
\end{align}
When both the target material and the FI are in thermal equilibrium, the lesser component of the dynamic spin susceptibilities satisfies
\begin{align}
    &G^<(\vk,\omega)=n_{\mathrm{BE}}(\omega)\qty[2i\mathrm{Im}G^{R}(\vk,\omega)], \\
    &\chi^<(\vq,\omega)=n_{\mathrm{BE}}(\omega)\qty[2i\mathrm{Im}\chi^{R}(\vq,\omega)],
\end{align}
where $n_{\mathrm{BE}}(\omega)=1/\qty(e^{\hbar\omega/k_{\mathrm{B}}T}-1)$ is the Bose-Einstein distribution function.
By employing the given expressions, we can confirm that the spin current vanishes in the thermal equilibrium.
In contrast, the lesser component of the dynamic spin susceptibility of the FI deviates from its thermal equilibrium value under the microwave irradiation because the magnons are excited.
Consequently, the spin current at the interface is given by
\begin{align}
    \la \IS\ra=
    &
    \frac{\hbar}{2}
    \int\frac{d\omega}{2\pi}
    \sum_{\vk,\vq}
    |J_{\vk,\vq}|^2
    \mathrm{Im}\chi^{R}(\vq,\omega)
    \mathrm{Im}
    \qty[-\delta G^<(\vk,\omega)],
\end{align}
where $\delta G^<(\vk,\omega)$ is the deviation from its thermal equilibrium.

\section{Collision terms}
\label{app:CollisionTerm}

Spin injection from the FI into the 2DEG is described by stochastic excitation induced by magnon absorption and emission. This process can be expressed by the collision term as
\begin{align}
& \frac{\partial f(\bm{k},\gamma)}{\partial t}\biggl{|}_{\rm pump} 
=\sum_{\bm{k}'}\sum_{\gamma'=\pm}
\Bigl{[}
P_{\bm{k}'\gamma'\rightarrow\bm{k}\gamma}
f(\bm{k}',\gamma')(1-f(\bm{k},\gamma))
-P_{\bm{k}\gamma\rightarrow\bm{k}'\gamma'}
f(\bm{k},\gamma)(1-f(\bm{k}',\gamma'))
\Bigl{]}, \label{eq:colpump1}
\end{align}
where $P_{\bm{k}\gamma\rightarrow\bm{k}'\gamma'}$ is the transition rate calculated with Fermi’s golden rule as
\begin{align}
P_{\bm{k}\gamma\rightarrow\bm{k}'\gamma'}
=\sum_{N_{\bm{0}}}\sum_{\Delta N_{\bm{0}} = \pm 1}
\frac{2\pi}{\hbar}
\Bigl{|}\langle \bm{k}',\gamma'|\langle N_{\bm{0}}+\Delta N_{\bm{0}} |{\cal H}_{\rm int}
|\bm{k},\gamma\rangle|N_{\bm{0}}\rangle \Bigl{|}^2\rho(N_{\bm{0}}) \delta\Bigl{(}E^{\gamma'}_{\bm{k}'}-E^{\gamma}_{\bm{k}}+\Delta N_{\bm{0}}\hbar\omega_{\bm{0}}\Bigl{)}, \label{eq:Pkkp}
\end{align}
where $|N_{\bm{0}}\rangle$ is the eigenstate of the magnon number operator, i.e., $b^{\dagger}_{\bm{0}}b_{\bm{0}}|N_{\bm{0}}\rangle=N_{\bm{0}}|N_{\bm{0}}\rangle$, $\Delta N_{\bm{0}}=\pm 1$ is a change of the magnon number, and $\rho(N_{\bm{0}})$ describes a nonequilibrium distribution function for the uniform spin precession driven by microwave. 
Assuming that the distribution function $\rho(N_{\bm{0}})$ has a sharp peak at its average $\langle N_{\bm{0}}\rangle$, the summation can then be approximated as $\sum_{N_{\bm{0}}} \rho(N_{\bm{0}}) F(N_{\bm{0}})
\simeq F(\langle N_{\bm{0}}\rangle)$, where $F(x)$ is an arbitrary function. 

The collision term due to impurity scattering is written as
\begin{align}
& \frac{\partial f(\bm{k},\gamma)}{\partial t}\biggl{|}_{\rm imp} =\sum_{\bm{k}'}\sum_{\gamma'=\pm}
\Bigl{[}
Q_{\bm{k}'\gamma'\rightarrow\bm{k}\gamma}
f(\bm{k}',\gamma')(1-f(\bm{k},\gamma))
-Q_{\bm{k}\gamma\rightarrow\bm{k}'\gamma'}
f(\bm{k},\gamma)(1-f(\bm{k}',\gamma'))
\Bigl{]}, \label{eq:colimp1}
\end{align}
where $Q_{\bm{k}\gamma\rightarrow\bm{k}'\gamma'}$ is the transition rate of electron scattering given as
\begin{align}
&Q_{\bm{k}\gamma\rightarrow\bm{k}'\gamma'} 
= \frac{2\pi}{\hbar}
\Bigl{|}\langle \bm{k}',\gamma'|H_{\rm imp}({\bm R})|\bm{k},\gamma\rangle \Bigl{|}^2 \delta\Bigl{(}E^{\gamma'}_{\bm{k}'}-E^{\gamma}_{\bm{k}}\Bigl{)}. \label{eq:Qkkp}
\end{align}
Note that the transition rates due to interfacial and impurity scattering include the overlap of the spin states between the initial and final states.

\bibliographystyle{iopart-num}
\bibliography{ref.bib}

\end{document}